\documentclass[a4paper,11pt]{article}
\usepackage[margin=1in]{geometry}
\usepackage[usenames,dvipsnames,table]{xcolor}
\usepackage{cite}
\usepackage{multirow}
\usepackage{xspace}
\usepackage{makecell}
\usepackage{pdflscape}
\usepackage{graphicx}
\usepackage{amssymb,amsmath,mathrsfs}
\usepackage{subfig}
\usepackage{authblk}
\usepackage[normalem]{ulem}
\usepackage[bookmarks,bookmarksnumbered,%
    citebordercolor={0.8 0.8 0.8},filebordercolor={0.8 0.8 0.8},%
    linkbordercolor={0.8 0.8 0.8},
    urlbordercolor={0.8 0.8 0.8}]{hyperref}

\usepackage{verbatim}
\usepackage{fancyhdr}

\renewcommand{\arraystretch}{1.5}

\newcommand{\Mpl}{M_\mathrm{Pl}}

\newcommand{\GeV}{\mathrm{GeV}}
\newcommand{\MeV}{\mathrm{MeV}}

\newcommand{\keV}{\mathrm{keV}}

\newcommand{\beq}{\begin{equation}}
\newcommand{\eeq}{\end{equation}}

\newcommand{\order}[1]{\mathcal{O}(#1)}

\DeclareMathOperator{\sech}{sech}
\DeclareMathOperator{\csch}{csch}
\newcommand{\Neff}{N_\mathrm{eff}}
\newcommand{\DNeff}{\Delta N_\mathrm{eff}}

\newcommand{\sds}{s_{\mathrm{ds}}}
\newcommand{\ssm}{s_{\mathrm{SM}}}
\newcommand{\rds}{\rho_{\mathrm{ds}}}

\newcommand{\rsm}{\rho_{\mathrm{SM}}}

\newcommand{\mds}{m_{\mathrm{ds}}}
\newcommand{\Tds}{T_{\mathrm{ds}}}
\newcommand{\Tsm}{T_{\mathrm{SM}}}
\newcommand{\TXeq}{T^{\ds\; \mathrm{eq}}}

\newcommand{\ds}{\mathrm{ds}}
\newcommand{\SM}{\mathrm{SM}}

\newcommand{\gcritds}{g_{\mathrm{critical}}^\ds}
\newcommand{\p}{\varphi}
\newcommand{\hc}{\mathrm{h.c.}}

\newcommand{\alterbbn}{\texttt{AlterBBN}\xspace}
\def\iso#1#2{\mbox{${}^{#2}{\rm #1}$}}
\def\he#1{\iso{He}{#1}}
\def\h#1{\iso{H}{#1}}
\def\li#1{\iso{Li}{#1}}
\def\be#1{\iso{Be}{#1}}
\def\D{\textrm{D}\xspace}
\def\H{\textrm{H}\xspace}
\def\cocetal{Coc \emph{et al.} (2015)\xspace}

\newcommand\FNAL{Fermi National Accelerator Laboratory, Batavia, IL, USA }
\newcommand\SLAC{SLAC National Accelerator Laboratory, 2575 Sand Hill Road, Menlo Park, CA, USA}
\begin{document}
\title{\bf Dark Sector Equilibration During Nucleosynthesis}
\author[1]{Asher Berlin}
\author[1,2]{Nikita Blinov}
\author[1]{Shirley Weishi Li}
\affil[1]{\SLAC}
\affil[2]{\FNAL}

\renewcommand{\headrulewidth}{0pt}
\setlength{\headheight}{40pt}
\fancypagestyle{plain}{
  \fancyhf{}
  \fancyhead[R]{ SLAC-PUB-17421 \\ FERMILAB-PUB-19-138-A-T}
}
\date{April 8, 2019}
\maketitle

\begin{abstract}
Light, weakly-coupled dark sectors may be naturally decoupled in the early universe and enter equilibrium with the Standard Model bath during the epoch of primordial nucleosynthesis. The equilibration and eventual decoupling of dark sector states modifies the expansion rate of the universe, which alters the predicted abundances of the light elements. This effect can be encompassed in a time-varying contribution to $N_\text{eff}$, the effective number of neutrino species, such that $N_\text{eff}$ during nucleosynthesis differs from its measured value at the time of recombination. We investigate the impact of such variations on the light element abundances with model-independent templates for the time-dependence of $N_\text{eff}$ as well as in specific models where a dark sector equilibrates with neutrinos or photons. We find that significant modifications of the expansion rate are consistent with the measured abundances of light nuclei, provided that they occur during specific periods of nucleosynthesis. In constraining concrete models, the relative importance of the cosmic microwave background and primordial nucleosynthesis is highly model-dependent.   
\end{abstract}

\tableofcontents

\section{Introduction}
Measurements of the light element abundances provide one of the earliest direct tests of cosmology.  Broad agreement of the observed abundances with predictions of standard Big Bang nucleosynthesis (BBN) has been used to constrain scenarios with late reheating~\cite{Kawasaki:2000en,Hannestad:2004px}, anti-matter domains~\cite{Rehm:1998nn}, long-lived particle decays~\cite{Fradette:2014sza,Fradette:2017sdd}, invisible~\cite{Hufnagel:2017dgo} and electromagnetic energy injection~\cite{Hufnagel:2018bjp,Forestell:2018txr,Depta:2019lbe}, and new light particles in thermal equilibrium with the Standard Model (SM)~\cite{Boehm:2012gr,Nollett:2013pwa,Nollett:2014lwa,Cyburt:2015mya}. For a survey of phenomena that can be tested by nucleosynthesis see, e.g., Ref.~\cite{Pospelov:2010hj}. Given the far-reaching implications of the observed primordial abundances of light nuclei, it is important to understand the robustness of the resulting conclusions.

In this work, we focus on possible modifications to the expansion rate at the time of primordial nucleosynthesis. Simple deformations of standard assumptions can leave open significant room for non-standard phenomena~\cite{Steigman:2013yua}. One key assumption that is often made when constraining light particles is that the new degrees of freedom were already in equilibrium with the SM at the onset of nucleosynthesis.  Light, weakly coupled dark sectors are typically \emph{not} in thermal equilibrium with the SM until late times~\cite{Bartlett:1990qq,Chacko:2003dt,Chacko:2004cz,Cadamuro:2010cz,Berlin:2017ftj,Berlin:2018ztp}. For instance, if equilibration takes place after neutrino-photon decoupling, the resulting modification to the expansion rate is suppressed~\cite{Chacko:2003dt,Chacko:2004cz,Berlin:2017ftj,Berlin:2018ztp}, allowing for the presence of many new degrees of freedom at the SM temperature.  Equilibration during nucleosynthesis is therefore a natural possibility.

\begin{figure}
  \centering
  \includegraphics[width=15cm]{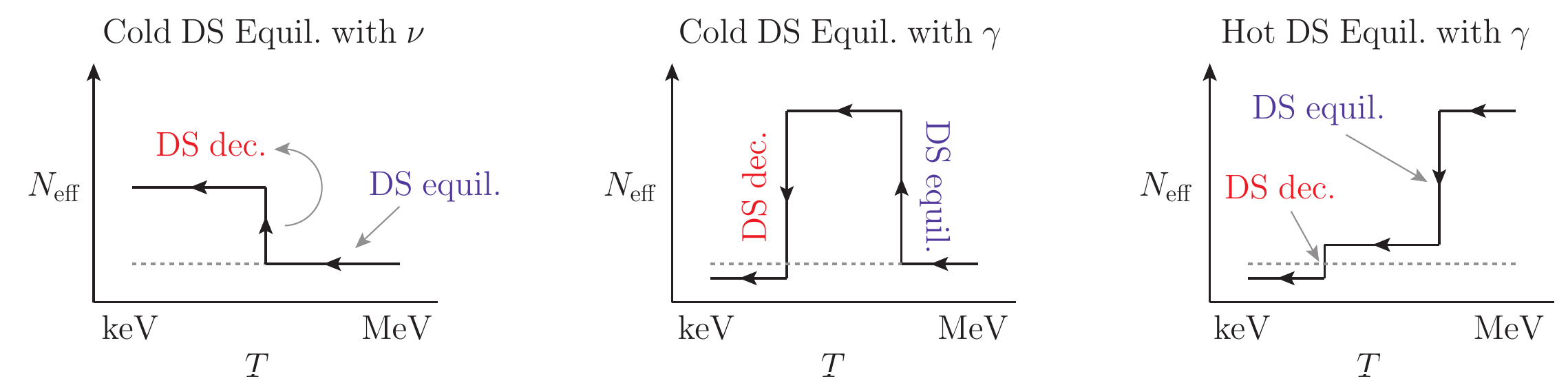}
  \caption{Schematic evolution of $\Neff$ for three representative cases: a dark sector (DS) equilibrates with neutrinos (left panel) or with photons (middle and right panels) after neutrino-photon decoupling. In the left panel, $\Neff$ does not change during equilibration due to energy conservation; it is only modified when the cold DS particles become non-relativistic and decouple, heating the neutrinos above the SM expectation (indicated by the gray dashed line).  In the middle panel, $\Neff$ increases when photons equilibrate with a cold DS. When these new particles decouple, they reheat the photon bath, leading to a decrease in $\Neff$. It is also possible that a DS that is initially hotter than the photon bath leads to a decrease in $\Neff$ during equilibration and decoupling, as shown in the right panel.  For the same number of DS degrees of freedom, the maximum deviation to $\Neff$ is larger for photon equilibration. See Sec.~\ref{sec:dark_sector_equilibration} for more details. 
  \label{fig:schematic_neff_evol}}
\end{figure}

The scenarios we consider are schematically shown in Fig.~\ref{fig:schematic_neff_evol}. At the onset of nucleosynthesis, the SM bath consists of two decoupled components -- photons and neutrinos. The equilibration of a light dark sector (DS) with one of these two SM components has a distinct impact on the effective number of neutrino species, $\Neff$, which parametrizes the energy density of the primordial plasma in non-electromagnetic degrees of freedom. New particles coupled to neutrinos (or those that are completely decoupled) increase $\Neff$ at late times, while those that interact with photons decrease $\Neff$.  Dark sector equilibration and decoupling with the SM can give rise to ``pulses'' in the modified expansion rate (and therefore $\Neff$) that occur during a specific stage of nucleosynthesis; this is illustrated in Fig.~\ref{fig:schematic_neff_evol}. Additionally, new species coupled to photons can dilute the baryon-to-photon ratio, $\eta_b$, which encodes the baryon asymmetry of the universe.

We discuss the conditions under which an initially decoupled DS attains equilibrium with the SM during nucleosynthesis in Sec.~\ref{sec:dark_sector_equilibration}. In Sec.~\ref{sec:bbn}, we summarize the sensitivity of the predictions of BBN to changes in the expansion rate and the baryon asymmetry. There, we also discuss our numerical methods and treatment of nuclear rate uncertainties.  We use this framework to first study the impact of time-dependent modifications to $\Neff$ in a model-independent manner, by considering generic templates for $\Neff$ variations that can be mapped onto specific models in Sec.~\ref{sec:modelindependent}.  We then analyze a concrete scenario where the leading interactions between the DS and SM are mediated by decays and inverse-decays of a light mediator in Sec.~\ref{sec:modeldependent}. In this case, we obtain a realistic time-evolution for $\Neff$ by solving the relevant Boltzmann equations for the DS and SM plasmas (with the relevant collision terms given in Appendix~\ref{sec:collision_terms}). We summarize our findings in Sec.~\ref{sec:conclusion}. A summary of common notation and definitions used throughout this work is provided in Table~\ref{Tab:notation}.

%%%%%%%%%%%%%%%%%%%%%%%%%%%%%%%%%%%%%%%%%%%%%%%%%%%%%%%%%%%%%%
%%%%%%%%%%%%%%%%%%%%%%%%%%%%%%%%%%%%%%%%%%%%%%%%%%%%%%%%%%%%%%
\section{Late Dark Sector Equilibration}
\label{sec:dark_sector_equilibration}

\subsection{Motivation}
\label{sec:motivation}

We focus on models in which a DS relativistically enters thermal equilibrium with the SM bath after neutrino-photon decoupling, which occurs at temperatures of $T \sim \text{few} \times \MeV$~\cite{Enqvist:1991gx,Dolgov:2002wy}. We will refer to this process as \emph{late equilibration}.\footnote{One can also call this \emph{late recoupling}, in analogy to \emph{re}combination; the prefix here does not necessarily imply that the DS and the SM were thermally coupled at earlier times.}  
These models are interesting for several reasons.
First, this cosmology is a generic feature of dark sectors with light (sub-MeV) feebly-interacting particles, as we argue below.
Second, strong constraints on thermalized sub-MeV particles derived from considerations of the cosmic microwave background (CMB) and BBN are significantly relaxed in these scenarios, as we will show in Sec.~\ref{sec:dark_sector_cosmo}. Finally, late equilibration provides an explicit example  
of a time-dependent modification of the Hubble expansion rate between nucleosynthesis and recombination, thereby 
highlighting the complementarity of BBN and CMB measurements in constraining the evolution of the early universe.

Late equilibration naturally occurs if a sub-MeV DS interacts with the SM through a light weakly-coupled force carrier. To see this, let us denote the rate for thermal equilibration between the DS and SM as $\Gamma_\text{eq}$. At temperatures much greater than the masses of various particles participating in the equilibrating reactions, the rate is schematically of the form
\beq
\Gamma_\text{eq} \sim \alpha_\ds^2 \, T ~~ (\text{DM-SM scattering})
\eeq
for DM-SM scattering and
\beq
\Gamma_\text{eq} \sim \alpha_\ds \, m_\ds^2 / T ~~ (\text{DS decays to SM})
\label{eq:gammaeq_decays}
\eeq
for decays and inverse-decays of DS force-carriers into the SM, where $T$ is the photon bath temperature, $\alpha_\ds$ is a coupling constant, and $m_\ds$ is a characteristic mass-scale in the DS. 
Equilibration occurs when $\Gamma_\text{eq}$ is comparable to the Hubble expansion rate,
\beq
H \sim T^2 / \Mpl ~,
\eeq
where $\Mpl$ is the Planck mass.  Since $\Gamma_\text{eq} / H$ increases as the universe cools, the DS will eventually enter equilibrium with the SM, provided that $\alpha_\ds$ is sufficiently large. Defining $T^{\ds \text{ eq}}$ to be the temperature at which $\Gamma_\text{eq} \sim H$, we find that the DS and SM enter equilibrium when the temperature drops below
\begin{align}
T^{\ds \text{ eq}} &\sim \alpha_\ds \, \Mpl ~~ (\text{scattering})
\nonumber \\
T^{\ds \text{ eq}} &\sim  \left( \alpha_\ds \, m_\ds^2 \, \Mpl \right)^{1/3} ~~ (\text{decays})
\label{eq:equilibration_temp}
~,
\end{align}
for scattering and decays, respectively. This occurs after neutrino-photon decoupling if the couplings are sufficiently small. For instance, decays and inverse-decays equilibrate the two sectors after neutrino-photon decoupling but before the end of nucleosynthesis if the DS coupling falls in the wide range, $\order{10^{-21}} \lesssim \alpha_\ds \, (m_\ds / \text{keV})^2 \lesssim \order{10^{-15}}$.

Once equilibrated, the DS will remain thermally coupled to the SM bath until the temperature drops below some characteristic mass scale (typically the mass of the heaviest particle involved in the reaction), at which point $\Gamma_\text{eq}$ becomes mass- or Boltzmann-suppressed and the DS decouples from the SM bath. 
As a result, each time the temperature drops below a DS mass-threshold, the corresponding species becomes non-relativistic and its comoving entropy density is transferred to the SM, similar to $e^+ e^-$ annihilations in the early universe. Throughout this work, we will refer to such processes as DS-SM \emph{decoupling}.

In this work, we consider two concrete examples in which a DS couples to neutrinos or photons. 
These scenarios affect the expansion rate of the universe in distinct ways. Late equilibration between a DS and the neutrino bath has been studied in Refs.~\cite{Chacko:2003dt, Chacko:2004cz} and 
was recently highlighted in Refs.~\cite{Berlin:2017ftj,Berlin:2018ztp} as a way of realizing cosmologies of thermal dark matter below the MeV scale.  
A generic feature of these models is a light mediator coupled to neutrinos.
Such particles are a natural feature of models of neutrino mass generation through spontaneous 
lepton-number ($L$) violation. The pseudo-Nambu-Goldstone boson of $L$-breaking, the majoron, is light and has 
renormalizable interactions with the neutrino mass-eigenstates. Moreover, the majoron can be coupled to additional 
light states that, for example, account for dark matter~\cite{Berlin:2017ftj,Berlin:2018ztp}.

Thus, the existence of a majoron can lead to the late equilibration of a DS with neutrinos.
In practice, we will consider a simplified model consisting of a light mediator, $\p$, interacting with neutrinos, 
\beq
\mathscr{L} \supset \lambda \, \p \, \nu^2 + \hc
~,
\label{eq:nu_interaction}
\eeq
where $\nu$ is a neutrino mass-eigenstate in two-component notation.
This coupling arises in a gauge-invariant manner after electroweak symmetry 
breaking mixes $SU(2)_L$-charged neutrinos with singlet right-handed neutrinos (with $\p$ directly coupled to the latter)~\cite{Berlin:2018ztp}.
The dominant process leading to energy transfer and equilibration between the SM and the DS 
is the decay $\p \leftrightarrow \nu\nu$. 

Late DS equilibration with photons was first considered in Ref.~\cite{Bartlett:1990qq} 
in a model with a millicharged particle and a dark photon. In that case, 
the dominant equilibrating reaction is a Compton-like scattering process.
For simplicity and in analogy to the neutrino case above, we will instead focus on a model where 
the dominant process is decay; this is realized by an axion-like particle, $\p$, with an interaction~\cite{Cadamuro:2010cz}
\beq
\mathscr{L}\supset \frac{\p}{4\Lambda} \widetilde{F}_{\mu\nu} F^{\mu\nu},
\label{eq:gamma_interaction}
\eeq
where $F$ is the photon 
field strength and $\Lambda$ encodes the scale 
and couplings of ultraviolet physics that generates this operator. 
As we show in Appendix~\ref{sec:photon_constraints}, in minimal models the size of $\Lambda$ that is required for equilibration to occur shortly before or during 
nucleosynthesis is typically ruled out by stellar cooling arguments and the observed SN1987A burst duration.
These bounds can potentially be avoided in more baroque scenarios~\cite{Berlin:2018ztp}. 
For simplicity, we will assume that the resolution of these issues has no impact on cosmology, 
especially on nucleosynthesis.

Since we are interested in the cosmological impact of late equilibration, we only use three parameters ($m_\text{ds}, g_\text{ds}, \Gamma_\text{eq}$ corresponding to the DS mass, DS internal degrees of freedom, and DS-SM equilibrating rate, respectively) 
to specify a model in which a DS couples to either neutrinos or photons.  In the next section, we will discuss in detail how this may lead to non-standard modifications to the expansion rate of the universe.

\renewcommand{\arraystretch}{1.5}
\begin{table}
\centering
\begin{tabular}{|c||c|c|}
\hline
\textbf{Notation} & \textbf{Definition} \\
\hline\hline
$T_i$ & temperature of species $i = \ds , \nu, \gamma$  \\ \hline
$T$ & shorthand for the photon temperature ($T_\gamma$)  \\ \hline
$\xi_i$ &  temperature of species $i$ normalized to the photon temperature  \\ \hline
$\xi_\ds^0$ &  value of $\xi_\ds$ before DS-SM equilibration  \\ \hline
$\xi_\nu^\text{SM}$ &  value of $\xi_\nu$ in a standard (unmodified) cosmology  \\ \hline
$T^{\text{ds eq}}$ & temperature of the photon bath at DS-SM equilibration  \\ \hline
\end{tabular}
\caption{Notation and various temperature scales discussed throughout this work.}
\label{Tab:notation}
\end{table}
%

%%%%%%%%%%%%%%%%%%%%%%%%%%%%%%%%%%%%%%%%%%%%%%%%%%%%%%%%%%%%%%
%%%%%%%%%%%%%%%%%%%%%%%%%%%%%%%%%%%%%%%%%%%%%%%%%%%%%%%%%%%%%%

\subsection{Impact on the Expansion Rate and the Baryon Density}
\label{sec:dark_sector_cosmo}

Light DS states can modify the history of the universe in several ways. 
The main effect considered in this work is on the Hubble expansion rate, which is determined by the total energy density, $H \propto \sqrt\rho$. 
As $\rho$ is dominated by relativistic species,  after neutrino-photon decoupling it receives contributions from photons, electrons, neutrinos, and potentially the DS:
\beq
\rho \approx \rho_\gamma + \rho_{e^\pm} + \rho_\nu + \rho_\ds
\, .
\eeq
The energy density of each relativistic species depends on its internal degrees of freedom, $g_i$, and its temperature, $T_i$. We define the effective relativistic degrees of freedom, $g_*^i$, such that $g_*^i = g_i$ for bosons and $g_*^i = (7/8) \, g_i$ for fermions. The energy density of each relativistic species is then given by $\rho_i = (\pi^2/30) \, g_*^i \, T_i^4$. 

It is convenient to parametrize the contribution of a light DS to the energy density in terms of 
the effective number of neutrino species, $\Neff$, which is defined as
\beq
\label{eq:totalrho}
\rho_\nu + \rho_\ds = \frac{7}{8} \left(\frac{\pi^2}{15}\right) \times \Neff(T) \times (T_\nu^\text{SM})^4
\, .
\eeq
Above, $T_\nu^\text{SM}$ is the neutrino temperature at a given time (or, equivalently, at a given photon 
temperature) in the standard cosmology and in the instantaneous-decoupling approximation. 
Under this approximation, $\Neff = 3$ in the SM.\footnote{Incomplete neutrino decoupling at the time of $e^+ e^-$ annihilation leads to 
neutrino spectral distortions and $\Neff \approx 3.045$~\cite{Mangano:2001iu,Mangano:2005cc,deSalas:2016ztq}. This subtlety is not 
accounted for in our treatment of early universe thermodynamics and nucleosynthesis, so we approximate $\Neff \approx 3$ in the SM.}
In scenarios with extra light degrees of freedom at the same temperature as the SM neutrinos, 
$\Neff$ simply counts these particles.
Below, we consider models in which the DS has a different initial temperature from that of the neutrinos, 
and the neutrino temperature evolution is modified; in this case, $\Neff$ no longer directly measures the number of particles in the DS. 
Equation~\ref{eq:totalrho} is consistent with the more common definition, $\Neff\propto (\rho_\nu + \rho_\ds)/\rho_\gamma$, 
often used in the context of the CMB. Our adopted definition of $\Neff$ also accounts for non-standard neutrino temperature evolution.

Equilibration or decoupling of any species after neutrino-photon decoupling typically results in time-evolution of $\Neff$. 
For example, in the SM, $e^\pm$ decoupling increases $\Neff$ to $\approx 3.045$~\cite{Mangano:2001iu,Mangano:2005cc,deSalas:2016ztq}.
Throughout this work, we use the photon temperature, $T \equiv T_\gamma$, as a standard reference for cosmological epochs. %clock.
For later convenience, we define $\xi_i$ as the relative temperature of species $i$ compared to that of the photon bath~\cite{Feng:2008mu},
\beq
\xi_i \equiv \frac{T_i}{T}
\, .
\eeq 
We use $\xi_\nu^{\mathrm{SM}}$ to denote the value of $\xi_\nu$ under the assumptions of a standard cosmology and instantaneous decoupling,
i.e., $\xi_\nu^{\mathrm{SM}}(T) = 1$ for $T\gg m_e$ and 
$\xi_\nu^{\mathrm{SM}}(T) = (4/11)^{1/3} \simeq 0.7$ for $T \ll m_e$~\cite{Kolb:1990vq}.
Equation~\ref{eq:totalrho} can be converted into an explicit expression for $\Neff$,
\beq
\label{eq:Neff0}
\Neff = 3 \bigg[ \left( \frac{\xi_\nu}{\xi_\nu^\text{SM}} \right)^4 + \frac{g_*^\ds}{g_*^\nu}  \left( \frac{\xi_\ds}{\xi_\nu^\text{SM}} \right)^4 \bigg]
\, ,
\eeq
where $g_*^\nu = 3 \times 2 \times (7/8) = 21/4$ and $g_*^\ds = g_*^\ds (T)$ is the (temperature-dependent) effective number of relativistic degrees of freedom in the DS. In general, $\xi_\nu$, $\xi_\nu^\text{SM}$, $\xi_\ds$, and $g_*^\ds$ all evolve with time (or temperature) as the universe expands and cools.

Dark sector equilibration is governed by the conservation of comoving energy. 
Consider, for example, a DS that equilibrates with neutrinos after they have themselves
decoupled from photons. 
 The energy of the closed DS+$\nu$ system is conserved up to the work done by the bath in driving the expansion of the universe.
As long as the participating DS and SM species are relativistic, energy conservation in an expanding universe takes the form
\beq
(\rsm + \rds)a^4 = \mathrm{constant}~~~\text{(SM--DS equilibration)} ,
\label{eq:sm_and_ds_energy_conservation1}
\eeq
where $a$ is the scale factor. The temperature evolution of the $\SM+\text{DS}$ bath can be estimated by equating 
the value of the left-hand side of Eq.~\ref{eq:sm_and_ds_energy_conservation1} immediately before and after equilibration  (assuming equilibration is instantaneous).  As this process is irreversible, entropy is not conserved.

As the universe continues to expand and cool, the temperature 
eventually drops below the masses of various DS degrees of freedom. As these DS particles become non-relativistic, they deposit their entropy into the $\SM+\text{DS}$ bath 
through, e.g., decays or annihilations and eventually decouple. This process occurs in equilibrium and therefore conserves entropy, 
leading to 
\beq
(\ssm + \sds) a^3 = \mathrm{constant}~~~\text{(SM--DS decoupling)}, 
\label{eq:sm_and_ds_entropy_conservation1}
\eeq
which can be used to estimate the $\SM + \text{DS}$ temperature before 
and after decoupling. Equations~\ref{eq:sm_and_ds_energy_conservation1}  and 
\ref{eq:sm_and_ds_entropy_conservation1} will be used to qualitatively
understand the temperature evolution of $\Neff$ for the scenarios 
considered below in Secs.~\ref{sec:neutrinocoupling} and \ref{sec:photoncoupling}.

Equilibration and decoupling are not instantaneous processes. The quantitative 
evolution of energy and entropy densities in the various sectors 
can be obtained from Boltzmann equations such as
\beq
\dot \rho_i + 3 H (\rho_i + p_i) = g_i \int \frac{d^3 p}{(2\pi)^3} C[f_j]
\, , 
\label{eq:energy_density_boltzmann1}
\eeq
where $i = \gamma, \nu, \ds$ and $C$ is the appropriate collision term.  
As discussed in Sec.~\ref{sec:motivation}, late equilibration of a feebly-interacting DS naturally occurs, for example, 
when the dominant process mediating DS-SM interactions is the decay of a light mediator, $\varphi$, into SM particles ($\varphi \to \SM$) 
with the decay rate, $\Gamma_\p$. The corresponding collision term is then given approximately by
\beq
g_i \int \frac{d^3 p}{(2\pi)^3} C[f_j] \simeq
- m_\varphi \Gamma_\varphi \left[n_\varphi(\Tds) - n_\varphi^\mathrm{eq}(\Tsm)\right],
\label{eq:simple_collision_term}
\eeq
where the superscript ``eq'' denotes an equilibrium distribution with $n_\varphi^\mathrm{eq}(T_i) \sim T_i^3$ in the relativistic limit. 
We will also assume the existence of number-changing processes in the DS that 
drive the DS (pseudo)-chemical potentials to zero; this 
simplification reduces the number of Boltzmann equations that need to be solved by allowing us to set $n_\varphi(\Tds) = n_\varphi^\mathrm{eq}(\Tds)$.
Pauli-blocking and Bose-enhancement factors modify the precise 
form of the collision term in Eq.~\ref{eq:simple_collision_term}, but their effects are always to establish thermal and chemical equilibrium 
between the DS and SM baths. We include quantum-statistical effects in numerical calculations in the model-specific results of 
Sec.~\ref{sec:modeldependent}; the corresponding collision terms are calculated in Appendix~\ref{sec:collision_terms}.

In the next two sections, we apply these results to particular scenarios where 
a light DS equilibrates with either neutrinos or photons.

\subsubsection{Coupling to the Neutrino Bath}
\label{sec:neutrinocoupling}

We use energy (Eq.~\ref{eq:sm_and_ds_energy_conservation1}) and entropy (Eq.~\ref{eq:sm_and_ds_entropy_conservation1}) conservation to understand 
the evolution of $\Neff$ throughout DS equilibration and decoupling. In the analytic expressions below, we take 
$T\lesssim m_e$ for simplicity, such that $g_*^\gamma = g_\gamma = 2$ and $\xi_\nu^{\mathrm{SM}} = (4/11)^{1/3}$.
Prior to equilibration, the DS contributes to the energy density like ``dark radiation'' such that 
\beq
\label{eq:Neff1}
\Neff \approx 3 + \frac{4}{7} \, g_*^\ds \left(\frac{\xi_\ds^0}{\xi_\nu^{\mathrm{SM}}}\right)^4,
\eeq
where $\xi_{\ds}^0$ is the DS-to-photon temperature ratio \emph{prior} to equilibration. 
Although we take $\xi_{\ds}^0$ to be a free parameter, one expects $\xi_{\ds}^0 < \xi_\nu^{\mathrm{SM}}$ if the DS has been diluted by entropy dumps (from SM particles decoupling or otherwise -- see, e.g.,~\cite{Brust:2013xpv,Patwardhan:2015kga}) 
or if the DS was not populated efficiently during reheating~\cite{Adshead:2016xxj}. 
It is also possible that the DS is initially hotter than the SM, but this scenario is generically in conflict with considerations of BBN and the CMB 
\emph{regardless} of when it attains equilibrium with neutrinos. This is easily seen 
from Eq.~\ref{eq:Neff1} since for $\xi_\ds^0/\xi_\nu^{\mathrm{SM}} > 1$ we have $\DNeff \gtrsim 0.6$. 
The subsequent dynamics of equilibration and decoupling from neutrinos only increase $\DNeff$ as described below, worsening 
the tension.\footnote{This is not obviously the case in the photon-equilibration scenario discussed below, since heating of the photons relative to neutrinos lowers $\Neff$.} 
For this reason, we focus on dark sectors that were initially colder than the SM neutrinos.
Note that this dark radiation contribution in Eq.~\ref{eq:Neff1} rapidly becomes negligible as $\xi_{\ds}^0$ is lowered. 

During DS-SM equilibration, the total comoving energy density is conserved (see Eq.~\ref{eq:sm_and_ds_energy_conservation1}). 
Since $\Neff$ is a measure of the total $\text{DS} + \nu$ energy density, it remains unchanged during this epoch. 
Therefore, immediately after DS-SM equilibration, the form of $\Neff$ is the same as in Eq.~\ref{eq:Neff1}. 
Note, however, that the neutrino and DS temperatures evolve non-trivially in order to conserve 
the comoving energy density (see, e.g., Fig.~1 of Ref.~\cite{Berlin:2017ftj}). 

As the universe cools, the temperature of the DS-$\nu$ bath may drop below 
a mass threshold of a DS particle. At this point, processes such as annihilations ($\text{DS}~ \text{DS} \to \nu \, \nu$) or decays ($\text{DS} \to \nu \, \nu$) efficiently deplete the thermal abundance of this DS particle, transferring   its entropy into neutrinos. 
This slightly increases the neutrino energy density relative to photons. 
To see this, let us assume that all $g_*^\ds$ degrees of freedom in the DS equilibrate with neutrinos after neutrino-photon decoupling and later simultaneously (and instantaneously) decouple. In that case, entropy conservation (Eq.~\ref{eq:sm_and_ds_entropy_conservation1})
implies that 
\beq
\Neff \approx 3\left(1 + \frac{g_*^\ds}{g_*^\nu}\right)^{1/3}
\left(1 + \frac{g_*^\ds}{g_*^\nu} \, (\xi_\ds^0)^4\right). 
\label{eq:neff_after_decoupling_nu_equil}
\eeq
For $\xi_\ds^0 < 1$ and $g_*^\ds \geq 1$, we therefore have $\Delta \Neff \gtrsim 0.18$. 
This result should be compared to the one obtained using the standard assumption 
that the DS species was in equilibrium with the SM before neutrino-photon decoupling. In this case, $\xi_\ds^0 = 1$ and 
 $\Delta \Neff$ is generically in conflict with considerations of BBN and the CMB 
even for $g_*^\ds = 1$~\cite{Cyburt:2015mya,Aghanim:2018eyx} .

\begin{figure}
  \centering
  \includegraphics[width=8cm]{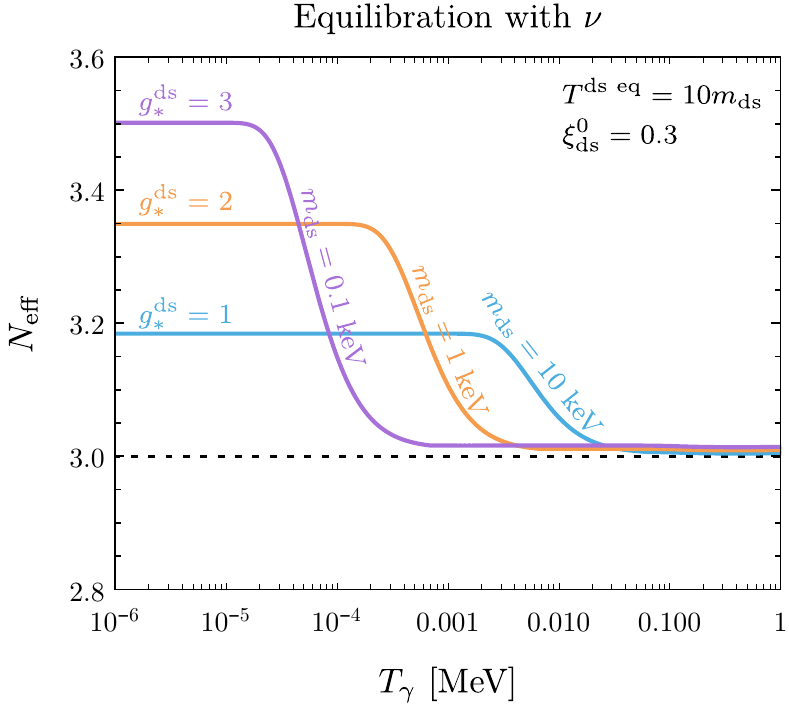}
  \caption{Evolution of $\Neff$ when a light DS equilibrates with neutrinos after neutrino-photon decoupling. The three curves 
    correspond to $(m_\ds, g_*^\ds) = (10\;\keV,1)$ (blue), $(1\;\keV,2)$ (orange), and $(0.1\;\keV,3)$ (purple), where $m_\ds$ is the common mass scale of the $g_*^\ds$ DS 
    degrees of freedom. For each curve, the initial DS-to-photon temperature ratio is fixed to $\xi_\ds^0 = 0.3$ and the SM-DS interaction rate is fixed to ensure equilibration at 
    $T^{\text{ds eq}} = 10 \, m_\ds$.
    Note that equilibration with neutrinos does not change 
    $\Neff$ because the total neutrino and DS energy density is conserved. $\Neff$ is only significantly modified when the DS degrees of freedom become non-relativistic and transfer their entropy to neutrinos when $T_\nu < m_\ds$. 
    \label{fig:neff_evol_nu}
    }
\end{figure}

The evolution of $\Neff$ for a DS equilibrating with neutrinos is shown schematically 
in the left panel of Fig.~\ref{fig:schematic_neff_evol}. Several detailed examples are 
also shown in Fig.~\ref{fig:neff_evol_nu}, for various representative values of DS degrees 
of freedom ($g_*^\ds$) at a common mass scale ($m_\ds$); 
these were calculated using the Boltzmann equation in Eq.~\ref{eq:energy_density_boltzmann1}. 
For each curve, the decay rate for DS $\leftrightarrow \nu\nu$ is chosen such that equilibration occurs at $T_\nu \approx 10 \, m_\ds$, 
fixing the initial DS-to-photon temperature ratio to be $\xi_\ds^0 = 0.3$.
Figure~\ref{fig:neff_evol_nu} illustrates that before and during equilibration, $\Neff$ is very close to 3. 
The DS mass scale, $m_\ds$, sets the decoupling temperature. 
After DS-$\nu$ decoupling, $\Neff$ is given by Eq.~\ref{eq:neff_after_decoupling_nu_equil}.  
If the various DS particles had parametrically different masses, these curves would contain 
additional ``steps'' corresponding to the different decoupling events as the temperature crosses these mass thresholds.
The value of $\Neff$ before and after 
each decoupling event can be estimated from repeated applications of comoving entropy 
conservation.

\subsubsection{Coupling to the Photon Bath}
\label{sec:photoncoupling}

Late equilibration of a DS with photons gives rise to a qualitatively different 
behavior of $\Neff$ compared to the neutrino case described above for two reasons.
First, after a new species equilibrates with the photon bath, a given photon temperature (which is effectively used as a ``clock" that determines the key cosmological epochs) corresponds to a 
\emph{larger} energy density (due to the increased degrees of freedom). 
This means that $\Neff$ changes during DS-SM equilibrium even though the comoving energy density is conserved during this process. 
Second, changes to the photon density modify the baryon-to-photon ratio, $\eta_b = (n_b-n_{\bar b})/n_\gamma$, 
which also alters the outcome of primordial nucleosynthesis.

As in the neutrino case, the DS contributes to $\Neff$ as dark radiation prior to equilibration such that 
\beq
\label{eq:Neff2a}
\Neff \approx 3 + \frac{4}{7} \, g_*^\ds \left(\frac{\xi_\ds^0}{\xi_\nu^{\mathrm{SM}}}\right)^4
\, ,
\eeq
where, as before, $\xi_{\ds}^0$ is the initial DS-to-photon temperature ratio.
DS-$\gamma$ equilibration fixes $\xi_{\ds} = 1$ and therefore 
modifies $\Neff$ to 
\beq
\label{eq:Neff2}
\Neff \approx 3 \left(\frac{g_*^\gamma + g_*^\ds}{ g_*^\gamma + g_*^\ds \, (\xi_{\ds}^0)^4} 
+ \frac{g_*^\ds / g_*^\nu}{(\xi_\nu^{\mathrm{SM}})^4}
\right)
\, .
\eeq
This equation exhibits two effects that modify $\Neff$ upon DS-$\gamma$ equilibration. The first term in Eq.~\ref{eq:Neff2} corresponds to $\xi_\nu^4$ in the general form of Eq.~\ref{eq:Neff0}. Depending on the initial temperature of the DS (encapsulated in $\xi_\ds^0$), DS-$\gamma$ equilibration either deposits energy into or draws 
energy from the $\gamma$ bath. As a result, the neutrino temperature (relative to photons) is lowered or increased, respectively.
The second term in Eq.~\ref{eq:Neff2} is the direct contribution from the thermalized DS component and simply counts the additional relativistic degrees of freedom. 

Both of these effects lead to non-trivial evolution of $\Neff$ that qualitatively depends on the size of the initial DS temperature ratio, $\xi_\ds^0$, and the number of DS degrees of freedom, $g_*^\ds$. 
Equation~\ref{eq:Neff2a} implies that $\Neff \gtrsim 3$ before equilibration. However, 
from Eq.~\ref{eq:Neff2} we see that after DS-$\gamma$ recoupling,
$\Neff \gtrsim 3$ if $g_*^\ds \gtrsim \gcritds$, or $\Neff \lesssim 3$ if $g_*^\ds \lesssim \gcritds$, 
where $\gcritds$ (the critical number of degrees of freedom in the DS) is given by
\beq
\label{eq:critical}
\gcritds \approx \frac{1}{(\xi_\ds^0)^4}  ~ \left[g_*^\nu \, (\xi_\nu^\text{SM})^4 \, \Big((\xi_\ds^0)^4 - 1\Big) - g_*^\gamma \right]
~.
\eeq
For an initially cold DS ($\xi_\ds^0 < 1$), $\gcritds$ is unphysical ($\gcritds < 0 < g_*^\ds$) and hence, $\Neff$ strictly increases during equilibration; this reflects the cooling 
of the photon bath relative to the neutrinos.

In principle, it is possible to realize a \emph{decrease} of $\Neff$ in a concrete %microphysical 
model.  We must demand that $\gcritds \gtrsim g_*^\ds > 1$, which is only possible if $\xi_\ds^0 \gtrsim 1.75$.
In this case, Eq.~\ref{eq:Neff2a} implies that $\Neff \gtrsim 24$ before equilibration.
Although such non-trivial behavior of $\Neff$ is interesting in this fine-tuned regime, this scenario is usually constrained since it necessarily leads to large deviations in the expansion rate at earlier times which will impact the decoupling of electroweak interactions.

\begin{figure}
  \centering
  \includegraphics[width=0.46\textwidth]{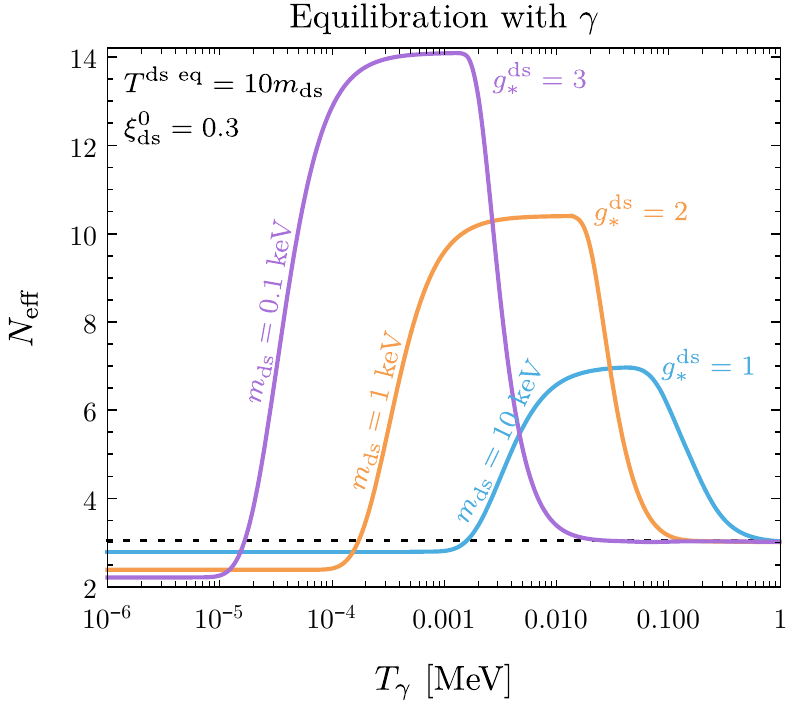}
  \includegraphics[width=0.46\textwidth]{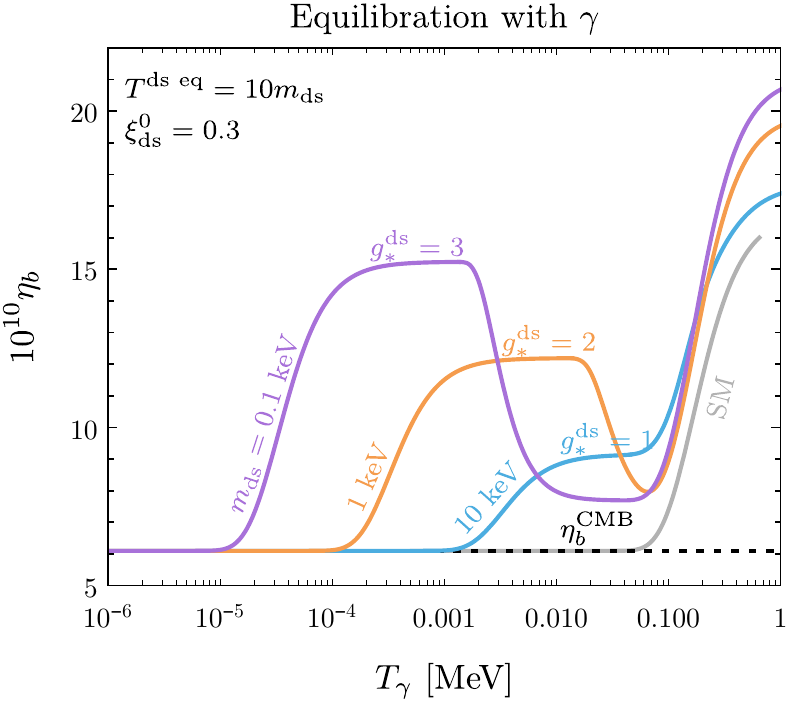}
  \caption{ 
As in Fig.~\ref{fig:neff_evol_nu}, but for a light DS that equilibrates with photons after neutrino-photon decoupling. 
In addition to $\Neff$ evolution (left panel), we also show the evolution of $\eta_b$ (right panel).
  Equilibration of the DS with photons lowers the photon temperature and introduces additional relativistic degrees of freedom into the plasma. 
    Both effects lead to a large increase of $\Neff$ at recoupling, while the lower photon temperature increases $\eta_b = (n_b - n_{\bar b})/n_\gamma$.
    When the DS particles become non-relativistic ($T\lesssim m_\ds$), the DS entropy is transferred back to the photons, reheating them slightly compared to neutrinos, resulting in $\Neff\lesssim 3$. In the right panel, we fixed the initial value of $\eta_b$ such that the baryon abundance matches the measured CMB value (indicated by the black dotted line) at late times. The light-gray solid line shows the evolution of $\eta_b$ in the SM, where the initial decrease at $T_\gamma \sim m_e$ 
    is due to entropy injection from $e^+e^-$ annihilations. 
    \label{fig:neff_and_etab_evol_gamma}}
\end{figure}

When the temperature of the DS-photon bath drops below a DS mass threshold, processes such as annihilations or decays transfer entropy to the photons, heating them relative to neutrinos and decreasing $\xi_\nu$. As a result, $\Neff$ decreases to 
\beq
\Neff \approx 3\left[ \Big(1 + \frac{g_*^\ds}{g_*^\gamma}\Big)^{1/3} \Big(1 + \frac{g_*^\ds}{g_*^\gamma} \, (\xi_\ds^0)^4\Big)\right]^{-1}
\, ,
\label{eq:neff_after_decoupling_photon_equil}
\eeq
which is strictly less than the SM expectation of $\Neff \approx 3$.
The evolution of $\Neff$ throughout DS-$\gamma$ equilibration and 
decoupling is illustrated schematically in the middle and right panels of Fig.~\ref{fig:schematic_neff_evol}, for an initially cold and hot DS, respectively.
In the left panel of Fig.~\ref{fig:neff_and_etab_evol_gamma} we show 
several examples of $\Neff(T)$ calculated using the Boltzmann equation in Eq.~\ref{eq:energy_density_boltzmann1}.

In addition to its effect on the expansion rate, DS-$\gamma$ equilibration modifies the predictions of BBN by changing the evolution of $\eta_b$. 
Suppose that before DS-$\gamma$ equilibration and $e^+ e^-$ annihilations, 
the baryon-to-photon ratio is given by $\eta_b^0$. Since $(n_b - n_{\bar b})/s_\nu$ is 
constant after neutrinos have decoupled, $\eta_b$ at a later time is 
given by 
\beq
\eta_b = \left(\xi_\nu\right)^3 \eta_b^0
\, .
\label{eq:etab_evolution}
\eeq
This parameter is independently determined by observations of the CMB to be $\eta_b^{\mathrm{CMB}} \approx 6.1\times 10^{-10}$ at the time of recombination~\cite{Aghanim:2018eyx}, fixing its value at late times. For a given cosmological history, which determines $\xi_\nu(T)$, the initial value of $\eta_b^0$ must be 
chosen such that the late-time value (Eq.~\ref{eq:etab_evolution}) is compatible with CMB measurements, i.e.,
\beq
\eta_b^0 = \xi_\nu(T_{\mathrm{CMB}})^{-3} \, \eta_b^{\mathrm{CMB}},
\label{eq:etab_initial_condition}
\eeq
where $T_\text{CMB} \sim 0.1 \text{ eV}$. In the case of DS-$\gamma$ equilibration, the evolution of the neutrino-to-photon temperature ratio, $\xi_\nu$, throughout equilibration and decoupling is given by
\beq
\xi_\nu \approx 
 \xi_\nu^\text{SM} \times \begin{cases}
 1 & (T > T^{\ds \text{ eq}} ) \\
 \left( \frac{g_*^\gamma + g_*^\ds}{g_*^\gamma + g_*^\ds \, (\xi_\ds^0)^4} \right)^{1/4} & ( m_\ds \lesssim T <  T^{\ds \text{ eq}}) \\
  \frac{\big( g_*^\gamma \big)^{1/3}}{\big(g_*^\gamma + g_*^\ds \big)^{1/12} \big(g_*^\gamma + g_*^\ds \, (\xi_\ds^0)^4 \big)^{1/4}} & (T \lesssim m_\ds)
  \, . 
\end{cases}
\eeq
Therefore, $\xi_\nu (T_\text{CMB})$ in Eq.~\ref{eq:etab_initial_condition} depends on when the DS equilibrates and decouples with respect to recombination.  The evolution of the baryon asymmetry (assuming the initial value given by Eq.~\ref{eq:etab_initial_condition}) is shown in the right panel of Fig.~\ref{fig:neff_and_etab_evol_gamma} for several choices of $g_*^\ds$ and $m_\ds$.

\section{Nucleosynthesis}
\label{sec:bbn}

The per-nucleon binding energies of nuclei are at the MeV scale. When the universe cools below this temperature, 
the formation of nuclei becomes energetically favorable. 
However, the large abundance of photons compared to baryons ensures that these bound states are not synthesized in significant numbers until much later, 
when the baryon plasma is even more dilute.
As a result, only \D, $^3$He, $^4$He, and $^7$Li are formed in appreciable quantities. Helium-4 and \D have the largest primordial abundances, which 
have been measured with several-percent precision. These observations are consistent with standard predictions of BBN, and in the following we use 
the abundances of these two elements to constrain beyond SM physics. In this section, we explain the effect of variations 
of $\Neff$ and $\eta_b$ on \he4 and \D yields; we also discuss the numerical treatment of nucleosynthesis for 
our modified cosmologies and derivation of model constraints.

We do not consider measurements of \he3 and \li7 in obtaining bounds on light dark sectors. There is no consensus on the measured \he3 primordial abundance due to limited observational regions and uncertainties in theoretical modeling~\cite{VangioniFlam:2002sa,Bania2007}.
The measured \li7 abundance is about a factor of three lower than the SM prediction, 
a long-standing discrepancy known as the ``Lithium Problem''~\cite{Fields:2011zzb}.
The \li7 abundance can be significantly modified after BBN in halo stars~\cite{Korn:2006tv}. 
Alternatively, the discrepancy can be due to physics beyond the SM~\cite{Goudelis:2015wpa,Jedamzik:2010lpi}.
Some of the scenarios considered in this work alleviate this tension, 
but at the price of making \he4 and \D yields inconsistent with observations. Throughout this work, we therefore 
assume that the resolution of the lithium problem is astrophysical in nature.

\subsection{Dependence on \texorpdfstring{$\Neff$}{Neff} and \texorpdfstring{$\eta_b$}{etab}}
\label{sec:dependence}

In Sec.~\ref{sec:dark_sector_cosmo}, we examined modifications to $\Neff$ and $\eta_b$ that arise as a result of a DS equilibrating and decoupling from either the neutrino or photon bath at temperatures below an MeV. How these (or any related) scenarios impact primordial nucleosynthesis can be understood in terms of a few key epochs that control \he4 and D yields: neutron-proton freeze-out, the deuterium bottleneck, and the freeze-out of deuterium burning processes -- 
see, e.g., Refs.~\cite{Esmailzadeh:1990hf,Smith:1992yy,Mukhanov:2003xs}. We now briefly review the essential physics for each of these epochs. 

\begin{figure}[t]
  \centering
  \includegraphics[width=0.6\textwidth]{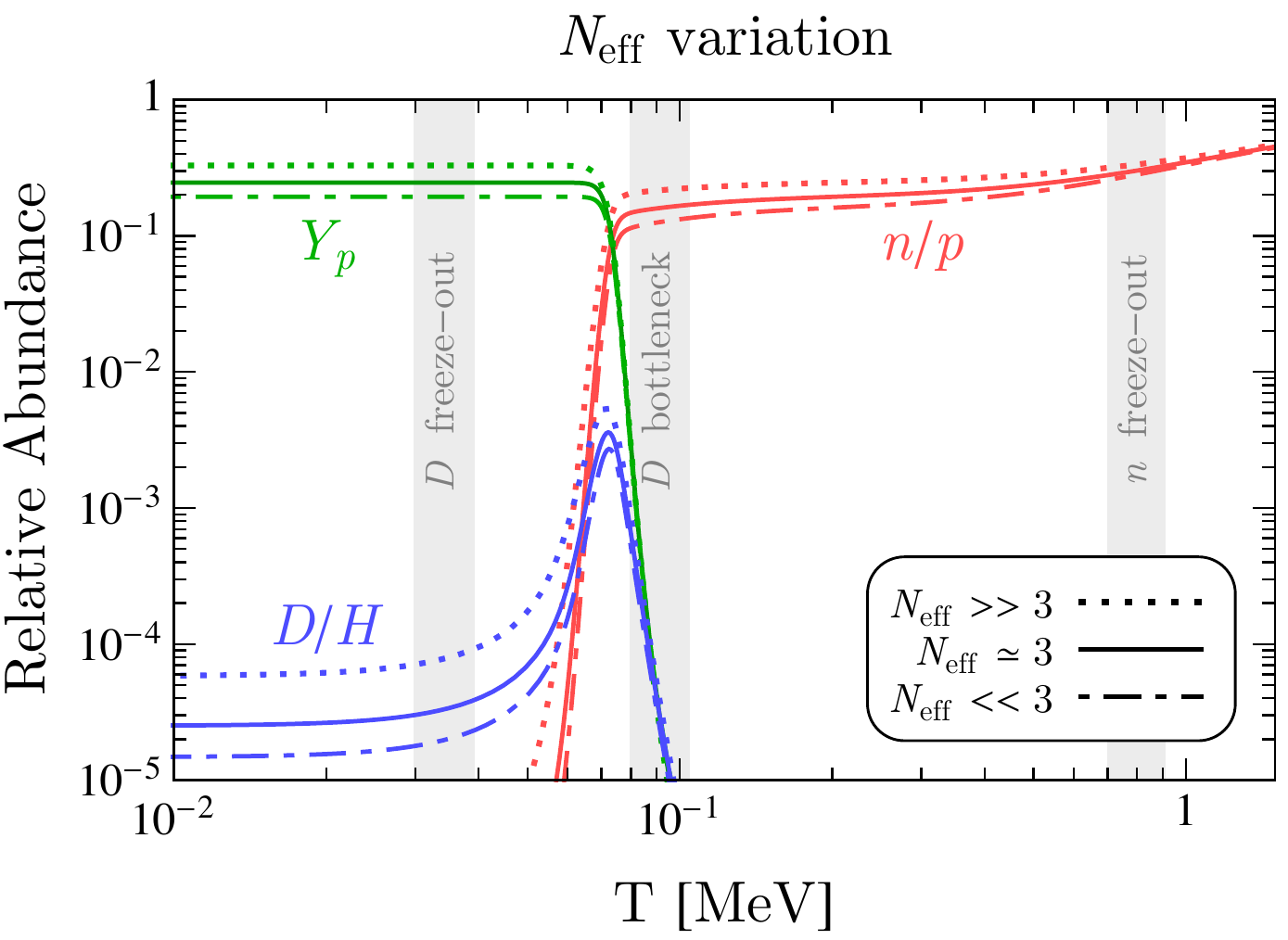}
  \caption{Effects of the expansion rate ($\Neff$) on the temperature evolution of the primordial abundances of neutrons ($n/p$) (red), helium-4 ($Y_p$) (green), and deuterium (D/H) (blue). The standard predictions of Big Bang nucleosynthesis are shown as the solid curves ($\Neff = 3$), while the dotted and dot-dashed lines correspond to $\Neff \gg 3$ ($\DNeff = 10$) and $\Neff \ll 3$ ($\DNeff = -3$). 
  \label{fig:BBN_evolution_standard_Neff}
} 
\end{figure}

In the early universe, neutrons and protons interconvert through electroweak interactions such as $n \, e^+ \leftrightarrow p \, \bar{\nu}_e$. At temperatures smaller than the neutron-proton mass-splitting ($T \lesssim \Delta m_{np} \approx 1.3 \text{ MeV}$), the neutron-to-proton number density ratio, $n/p$, becomes smaller than unity, i.e., $n / p \approx \text{exp}(-\Delta m_{np}/T) \lesssim 1$. Eventually, neutron-proton conversion processes freeze-out at $T \approx 0.8 \text{ MeV}$ at which point $n/ p \approx \text{exp}(-1.3/0.8) \approx 1/5$.\footnote{In reality, neutron-proton freeze-out is not instantaneous, but occurs over a range of temperatures. For more details, see, e.g. Ref.~\cite{Grohs:2016vef}.}
The decays of free neutrons further reduces the neutron density to $n/p \approx 1/7$ before the onset of nucleosynthesis.
Nearly all of these neutrons are eventually bound into \he4 because of its large binding energy and the corresponding 
enhanced equilibrium number density.
The resulting \he4 mass-fraction, $Y_p$, is therefore given by
\beq
\label{eq:Ypestimate}
Y_p \approx \frac{4 \, n_\text{He}}{n_p + n_n} \approx \frac{4 \, (n_n/2)}{n_p + n_n} = \frac{2 \, (n/p)}{1+(n/p)}  \approx \frac{1}{4}
~.
\eeq
Although the per-nucleon binding energies of light nuclei are $\mathcal{O}(\text{MeV})$, nucleosynthesis  
does not occur until much smaller temperatures near $T \sim 100 \text{ keV}$.
There are two reasons for this: the large entropy of the universe and the small deuterium binding energy.
The first fact implies that even if, e.g., \he4 was in chemical equilibrium with $n$ and $p$ its 
abundance would be negligible until $T\sim 300\;\keV$~\cite{Kolb:1990vq}.
The synthesis of these heavier nuclei is delayed even further because their 
production rates, e.g., via $\D \; \D \to p \; \h3$ (the first step in making \he4), depend on the deuterium abundance.
These reaction rates are slow compared to Hubble expansion until $n_{\D}/{(n_p + n_n)} \sim \order{10^{-4}}$~\cite{Mukhanov:2003xs}.
This rate-limiting step is often referred to as the ``deuterium bottleneck."
The temperature at which the \D-burning reactions become important 
can be estimated from the relative equilibrium abundance,
\beq
\label{eq:bottleneck}
\frac{n_{\rm D}}{n_p + n_n} \sim \eta_b \left( \frac{T}{m_p} \right)^{3/2} e^{B_{\rm D} / T}
~,
\eeq
where $B_{\rm D} \sim 2 \text{ MeV}$ is the deuterium binding energy. This ratio is initially $\ll 1$ because of the large entropy of the plasma (small baryon-per-photon number, $\eta_b \sim 10^{-10}$). 
This abundance becomes large enough to support the production of heavier nuclei only when the temperature falls below $\sim 100\;\keV$.
At this point, nucleosynthesis commences, and most of the deuterium is eventually processed into \he4. 
However, deuterium-burning processes (such as $\D \; p \to \he3 \; \gamma$) remain effective until they freeze out at temperatures of $T \sim \text{few} \times 10 \text{ keV}$ and the deuterium abundance has been depleted down to trace amounts of $\D/\H \sim 10^{-5} - 10^{-4}$.

\begin{figure}[t]
  \centering
  \includegraphics[width=0.6\textwidth]{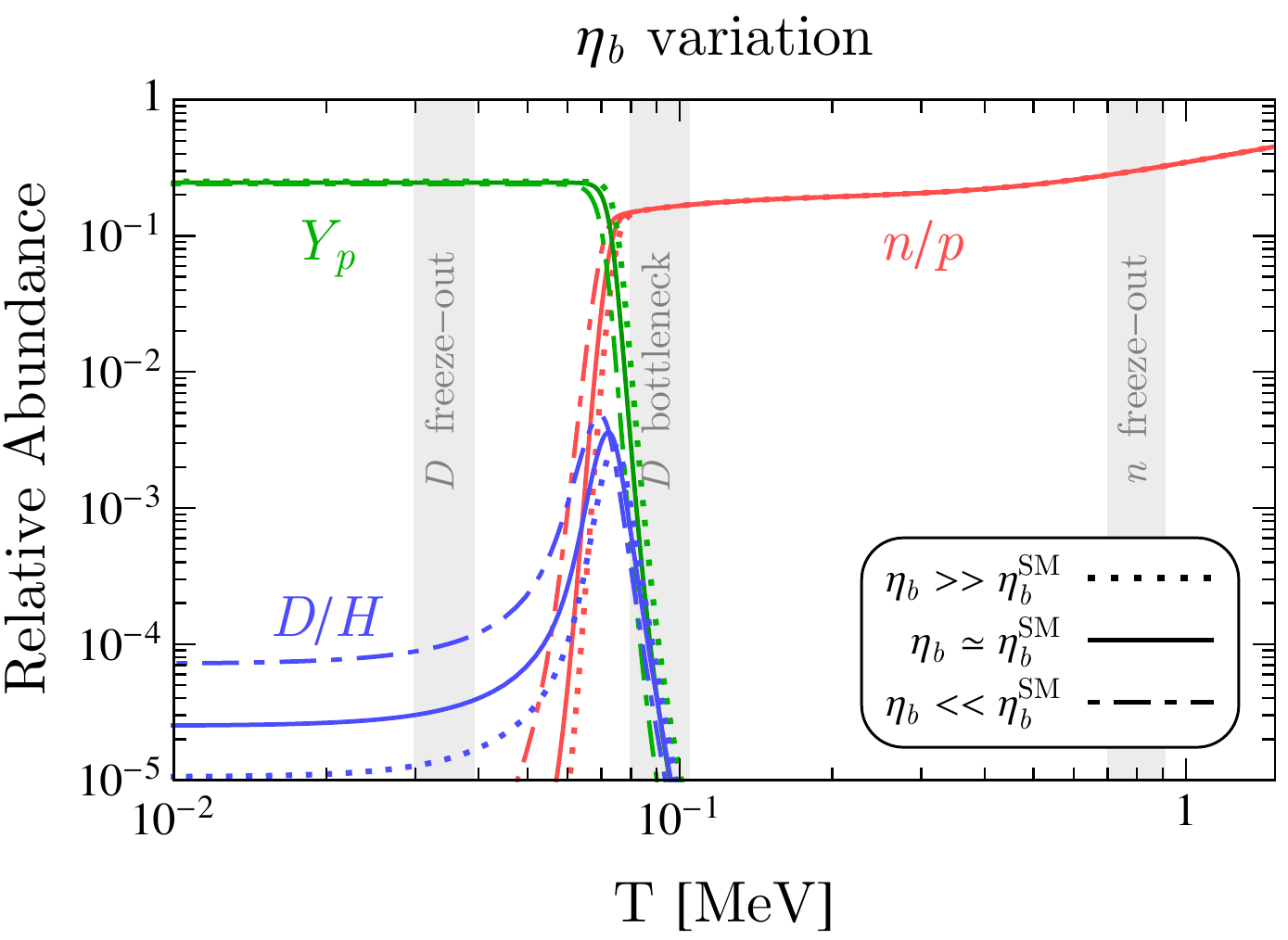}
  \caption{Effects of the baryon density ($\eta_b$) on the temperature evolution of the primordial abundances of neutrons ($n/p$) (red), helium-4 ($Y_p$) (green), and deuterium (\D/\H) (blue). The standard predictions of Big Bang nucleosynthesis are shown as the solid curves ($\Delta \eta_b = 0$), while the dotted and dot-dashed lines correspond to $\eta_b \gg \eta_b^\text{SM}$ ($\Delta \eta_b = 4 \times 10^{-10}$) and $\eta_b \ll \eta_b^\text{SM}$ ($\Delta \eta_b = - 3 \times 10^{-10}$).
  \label{fig:BBN_evolution_standard_eta}
} 
\end{figure}

We show how theoretical predictions of the neutron-to-proton ratio ($n/p$), the \he4 mass-fraction ($Y_p$), and the deuterium fraction (D/H) are affected by modifications to the expansion rate in Fig.~\ref{fig:BBN_evolution_standard_Neff} and the baryon density in Fig.~\ref{fig:BBN_evolution_standard_eta}, encapsulated in $\Neff$ and $\eta_b$, respectively. 
Values of $\Neff$ that are greater or smaller than the SM prediction of $\Neff \approx 3$, correspond to an increased or decreased expansion rate during nucleosynthesis. For $\Neff \gtrsim 3$, neutron-proton interconverting processes decouple at earlier times (compared to the SM prediction), leading to an increase in the predicted value of $n/p$ and hence a larger \he4 yield (see Eq.~\ref{eq:Ypestimate}). Also for $\Neff \gtrsim 3$, D-burning reactions freeze out earlier in the more rapidly expanding universe, which results in a larger predicted deuterium abundance. Conversely, $Y_p$ and D/H decrease for $\Neff \lesssim 3$. 

The \he4 mass-fraction is only logarithmically sensitive to the baryon density, $\eta_b$. This effect arises due to how $\eta_b$ controls the deuterium bottleneck. Defining the temperature at which nucleosynthesis commences as $T_\text{BBN} \sim \order{100} \text{ keV}$ through the criterion $n_{\rm D} / (n_p + n_n) \sim \order{10^{-4}}$~\cite{Mukhanov:2003xs}, we see from Eq.~\ref{eq:bottleneck} that
\beq
T_\text{BBN} \sim B_{\rm D} / \log{(1/\eta_b)} \sim \order{100} \text{ keV} \times \left(  1 + \order{10^{-2}} \, \frac{\delta \eta_b}{\eta_b} \right)
~,
\eeq
where in the second equality, we have considered small variations of $\eta_b$ around the CMB-preferred value, i.e., $\eta_b \sim \order{10^{-10}} + \delta \eta_b$. Hence, for much larger $\eta_b$, nucleosynthesis begins at earlier times, and less unbound neutrons decay before ending up in \he4. As a result,  $Y_p$ increases for larger $\eta_b$. The onset of nucleosynthesis corresponds to times much shorter than the neutron lifetime, such that $Y_p$ is only power-law (not exponentially) dependent on $T_\text{BBN}$. Unlike \he4, deuterium is much more sensitive to deviations in the baryon density, since $\eta_b$ directly controls the strength of D-burning rates. In particular, larger $\eta_b$ corresponds to more efficient D-burning and a smaller trace abundance at late times.

\subsection{Procedure}
\label{sec:procedure}

The primordial \he4 and D densities are inferred from direct observations of various astrophysical sites. For instance, the primordial \he4 abundance is measured from observations of recombination emission lines originating from low-metallicity HII regions of galaxies~\cite{Izotov:2014fga,Aver:2015iza,Peimbert:2016bdg}. The observed deuterium abundance constitutes a lower bound on its primordial value since it is easily destroyed in stellar cycles. It is directly observed from, e.g., isotope-shifted $\text{Ly}\alpha$ features in the absorption spectra of distant quasars~\cite{Cooke:2016rky,Balashev:2015hoe}. In this work, we adopt the recommended values of Ref.~\cite{PhysRevD.98.030001} for the observed abundances,
\beq
\label{eq:obsyields}
(Y_p)^\text{obs.} = 0.245 \pm 0.003~,~~ (\D/\H)^\text{obs.} = (2.569 \pm 0.027) \times 10^{-5}
~.
\eeq

We calculate \he4 and \D yields by modifying version 1.4 of the publicly available code \alterbbn~\cite{Arbey:2011nf}.  Similar to many public BBN programs (see, e.g., Refs.~\cite{Consiglio:2017pot,Pitrou:2018cgg}), \alterbbn follows the structure and techniques of the seminal Kawano code~\cite{Kawano:1992ua}. These codes take cosmological parameters and tabulated nuclear reaction rates as inputs and estimate the primordial abundances by solving a set of differential equations governing BBN.  For more details, see Refs.~\cite{Kawano:1992ua, 1969ApJS...18..247W, Jenssen2016}. 

The default version of \alterbbn computes primordial abundances assuming a standard cosmological history. In Sec.~\ref{sec:dark_sector_equilibration}, we discussed how light dark sectors that equilibrate during nucleosynthesis lead to temperature-dependent modifications to the expansion rate and the baryon density. We have modified \alterbbn in order to calculate the primordial nuclei yields for such general temperature-dependent forms of $\Neff (T)$ and $\eta_b (T)$.\footnote{Throughout this work, we assume that the temperature of neutrino-photon decoupling ($\sim \text{few} \times \text{MeV}$) is unchanged.}

The particular values of nuclear reaction rates and their experimental errors have a large impact on the predicted primordial abundances and their theoretical uncertainties.  The default rates used in \alterbbn are listed in Table~1 of Ref.~\cite{Arbey:2011nf}. We have updated the nuclear reactions that are relevant to \he4 and \D with newer calculations, largely following the choices of Ref.~\cite{Cyburt:2015mya}. The modified rates are summarized in Table~\ref{tab:nuclear_rates}. We mostly utilize results from the NACRE collaboration, which compiles and evaluates the latest updated cross sections, as given in the NACRE-II compilation~\cite{Xu:2013fha}.  However, two important reactions are not provided in the NACRE-II compilation: $p \; n \to \D \; \gamma$ and $\he3 \; n \to p \; \h3$. For these, we use calculations from Refs.~\cite{Ando:2005cz, Cyburt:2004cq}, respectively. We note that various reactions rates have been updated since the NACRE-II release.  In particular, Ref.~\cite{Coc:2015bhi} updated 
three important reactions of D burning.  Not only do they shift the central values of these rates by $\sim 10$\%, they also report much smaller errors.  Both are crucial for constraining new physics scenarios with observations of deuterium.  In our baseline analysis, we follow Ref.~\cite{Coc:2015bhi} for these three rates because the methodology used to calculate the central values and their spreads is comprehensively documented, with special attention paid to experimental systematic errors and theoretically motivated fitting functions. In Sections~\ref{sec:standard} and \ref{sec:modeldependent}, we compare results for different choices of \D-burning rates. 
In the models we study in this work, the neutrino temperature is modified compared to the SM, which changes the $p\leftrightarrow n$ conversion rate. 
We have checked that varying the neutrino temperature by a factor of a few negligibly affects the conversion rate and we therefore ignore this effect.

\begin{table}
  \centering
  \begin{tabular}{|c | c|}
    \hline
    Rate & Reference\\
    \hline
    $p (n,\gamma)\D$ & \cite{Ando:2005cz} \\
    $\he3(n,p)\h3$  & \cite{Cyburt:2004cq} \\
    \hline
    $\D(\D,\gamma)\he4$     & \multirow{8}{*}{\cite{Xu:2013fha}}\\
    $\h3(\D,n)\he4$         &   \\
    $\he3(\D,p)\he4$        &   \\
    $\D(\he4 ,\gamma )\li6$ &   \\
    $\li6(p,\gamma )\be7$   &   \\
    $\li7(p,\gamma )\be8$   &   \\
    $\li6(p,\he4 )\he3$     &   \\
    $\li7(p,\he4 )\he4$     &   \\
    \hline           
    $\D(\D,n)\he3$ &  \multirow{3}{*}{\cite{Xu:2013fha,Coc:2015bhi}}  \\
    $\D(p,\gamma)\he3$ & \\ 
    $\D(\D,p)\h3$ & \\
    \hline
  \end{tabular}
  \caption{Nuclear rates updated relative to \alterbbn v1.4~\cite{Arbey:2011nf}.
  In the text, we compare the predictions using NACRE-II~\cite{Xu:2013fha} and 
  \cocetal~\cite{Coc:2015bhi} for the reactions in the last three rows due to their 
  importance in estimating the deuterium yield.
  \label{tab:nuclear_rates}}
\end{table}

Theoretical uncertainties of the predicted \he4 and \D yields stem from uncertainties in various nuclear reaction rates and the neutron lifetime.  
We determine the 
corresponding error bars from a Monte Carlo procedure similar to 
that outlined in Ref.~\cite{Cyburt:2015mya}. We estimate the theoretical uncertainties 
for \he4 and \D yields 
by lognormal-sampling nuclear rates~\cite{Longland:2010gs,Iliadis:2010mj} and computing the yields $10^4$ times for each 
value of $\Neff$ and $\eta_b$.
The resulting yield distribution is well-described by a correlated Gaussian likelihood. When we utilize 
the \D-burning rates from the NACRE-II compilation (see the last three rows in Table~\ref{tab:nuclear_rates}), 
we obtain uncertainties similar to those of Ref.~\cite{Cyburt:2015mya}: 
\begin{align}
&\text{\underline{NACRE-II}}
\nonumber \\
& Y_p = 0.24633 \pm 0.00034 \, , \, 10^5 \, \mathrm{D/H} = 2.57\pm 0.12 ~~(10^{10} \, \eta_b^\text{CMB} = 6.10) % fixed eta = 6.10
 \nonumber \\
& Y_p = 0.24633 \pm 0.00034 \, , \, 10^5 \, \mathrm{D/H} = 2.57\pm 0.13 ~~(10^{10} \, \eta_b^\text{CMB} = 6.10 \pm 0.04 ) % vary eta +/- 0.04
\, , 
\label{eq:nacre_uncertainties}
\end{align}
where we have taken the SM expectation of $\Neff \approx 3$, in the first line we have fixed the late-time baryon density to the central value inferred from observations of the CMB, and in the second line we have additionally incorporated the spread around the central value, $\eta_b^\text{CMB} = (6.10 \pm 0.04) \times 10^{-10}$~\cite{Cyburt:2015mya,Ade:2015xua}.  In the first and second lines of Eq.~\ref{eq:nacre_uncertainties}, we find a correlation coefficient of $\approx -0.22$ and $\approx -0.26$, respectively.
The anti-correlation is driven primarily by the 
sizable spread in $\D \; \D \to n \; \he3$ and $\D \; \D \to p \; \h3$ (larger values of these rates reduce D/H and enhance \he4 abundances~\cite{Cyburt:2015mya}).

Compared to the NACRE-II compilation, the uncertainties of important \D-burning processes (last three rows of Table~\ref{tab:nuclear_rates}) 
are much smaller in Ref.~\cite{Coc:2015bhi}. If we instead adopt these rates from Ref.~\cite{Coc:2015bhi}, we find that for $\Neff \approx 3$,
\begin{align}
&\text{\underline{\cocetal}}
\nonumber \\
&Y_p =  0.24642 \pm 0.00032 \, , \, 10^5 \, \mathrm{D/H} = 2.446\pm 0.037 ~~(10^{10} \, \eta_b^\text{CMB} = 6.10) % fixed eta = 6.10
\nonumber \\
&Y_p =  0.24642 \pm 0.00033 \, , \,  10^5 \, \mathrm{D/H} = 2.446\pm 0.046 ~~(10^{10} \, \eta_b^\text{CMB} = 6.10 \pm 0.04) % vary eta +/- 0.04
\, ,
\label{eq:cocetal_uncertainties}
\end{align}
where $\eta_b$ is fixed as in Eq.~\ref{eq:nacre_uncertainties}. We find a correlation coefficient of $\approx 0.006$ and $\approx -0.10$ in the first 
and second lines, respectively; the correlation is mostly due to the $\eta_b$ sensitivity of the two yields, as discussed in Sec.~\ref{sec:dependence}. 
The central value for the predicted deuterium abundance in Eq.~\ref{eq:cocetal_uncertainties} is slightly smaller than that of Eq.~\ref{eq:nacre_uncertainties}. This shift is due to the larger D-burning rates of Ref.~\cite{Coc:2015bhi}. 

The theoretical uncertainty of the \he4 abundance is negligible compared to the observational one.  This is because the \he4 abundance is mainly sensitive to the neutron-to-proton ratio.
Therefore, the dominant theoretical uncertainty for $Y_p$ comes from  
variations of the neutron lifetime, which we take from Ref.~\cite{PhysRevD.98.030001}:
\beq
\tau_n = 880.2\pm 1.0\;\mathrm{s}.
\label{eq:neutron_lifetime}
\eeq
There is a well-established tension between bottle and beam measurements of the 
neutron lifetime~\cite{Wietfeldt:2011suo}. The value quoted above is an average that is dominantly determined by bottle experiment measurements. We note that even if the true value of $\tau_n$ is closer to that inferred from beam-based measurements ($\sim 888$ s), the corresponding shift to $Y_p$ and $\D/\H$ would be at the sub-percent level.

We find that the fractional theoretical uncertainties, $\sigma(X)/X$ for $X=Y_p$ and $\D/\H$, are 
to a good approximation independent of the functional form of $\Neff(T)$ and $\eta_b(T)$. 
This can be understood as follows. 
  Reference~\cite{Fiorentini:1998fv} has shown that the nuclear rate uncertainties in the final 
  yields are well-described using linear error propagation. As a result, the sensitivity of 
   yields of element $X$ to the rates, $\Gamma_i$, is encompassed by a set of 
  constants, $\alpha_i$ (the logarithmic derivatives of the yields with respect to the rates), such that
  \beq
  X \propto \prod_i {\Gamma_i}^{\alpha_i}.
  \eeq
In the nucleosynthesis Boltzmann system, these 
rates enter as $\Gamma_i/H$, which suggests that the effect of variations of $H$ (via $\Neff$) can be 
likewise linearized, at least for small deformations of $H$, leading to 
$X \propto {\Neff}^{\alpha^\prime}$, for some other constant $\alpha'$.
The resulting set of logarithmic derivatives, $\{\alpha_i,\alpha'\}$, is presented in Ref.~\cite{Cyburt:2015mya}.
A similar argument can be made for modifications to $\eta_b$.
Using this linearized form of the yields, it is simple to show that fractional errors from nuclear 
rate uncertainties, $\sigma(X)/X$, are constant as  
$\Neff$ and $\eta_b$ are varied. This is expected to hold for small perturbations 
to these quantities (as long as the linearization is valid); fortunately, 
the measured and predicted abundances of the light elements are so precise that only 
small deviations from standard values are allowed. 
While these arguments are straightforward for constant shifts to $\Neff$ and $\eta_b$, 
they are more difficult to make for arbitrary time-variations of these quantities.
Therefore, we have explicitly checked the constancy of $\sigma(Y_p)/Y_p$ and $\sigma (\D/\H) / (\D/\H)$
for time-dependent variations of $\Neff$ and $\eta_b$, as considered in the following sections, 
using the Monte Carlo approach described above. 
This justifies our use of $\sigma(Y_p)/Y_p$ and $\sigma (\D/\H) / (\D/\H)$ from 
Eqs.~\ref{eq:nacre_uncertainties} and~\ref{eq:cocetal_uncertainties} throughout this work. 
Although we adopt Eq.~\ref{eq:cocetal_uncertainties} for our baseline analysis, in Secs.~\ref{sec:standard} 
and~\ref{sec:modeldependent} we show how these two rate choices affect the constraints for constant 
shifts to $\Neff$ and $\eta_b$ and in concrete particle physics models, respectively.

\subsection{Standard Constraints on \texorpdfstring{$\Neff$}{Neff} and \texorpdfstring{$\eta_b$}{etab}}
\label{sec:standard}

Before we discuss our main results in Sec.~\ref{sec:results}, we first present a simple estimate for the standard bounds on $\Neff$ and $\eta_b$, assuming that they take constant values throughout the epoch of primordial nucleosynthesis. 
These results exemplify our methods that we will apply again in Sec.~\ref{sec:results}. 

We utilize the methodology outlined in Sec.~\ref{sec:procedure} in order to calculate \he4 and D yields, $(Y_p)^\text{theory}$ and $(\D/\H)^\text{theory}$, respectively. We then compare these predictions to the observed abundances in Eq.~\ref{eq:obsyields} through the $\chi^2$ test statistic,
\beq
\chi^2 = (\Delta Y_p)^2 + (\Delta \D/\H)^2
~,
\label{eq:chi2_def}
\eeq
where we have defined
\beq
\Delta Y_p \equiv \frac{ \, (Y_p)^\text{theory} - (Y_p)^\text{obs.}}{\Big[ (\sigma_{Y_p}^\text{theory})^2 + (\sigma_{Y_p}^\text{obs.})^2 \Big]^{1/2}}~~,~ ~
\Delta \D/\H \equiv \frac{ \, (\D/\H)^\text{theory} - (\D/\H)^\text{obs.}}{\Big[ (\sigma_{\D/\H}^\text{theory})^2 + (\sigma_{\D/\H}^\text{obs.})^2 \Big]^{1/2}}
~~.
\label{eq:delta_def}
\eeq
Above, $\sigma_{Y_p, \D/\H}^\text{obs.}$ and $\sigma_{Y_p, \D/\H}^\text{theory}$ are the observational and theoretical uncertainties for $Y_p$ and $\D/\H$, as detailed in Sec.~\ref{sec:procedure}. 
Fixing $\Neff$ and $\eta_b$ to their values expected in a standard cosmology, we find that
\beq
\label{eq:SMDelta}
\Delta Y_p \approx 0.46 ~~,~~ \Delta \D / \H \approx -2.34 ~~\text{(standard cosmology)}
~.
\eeq
This gives $\chi^2 \approx 5.7$, assuming a standard cosmological history.

\begin{figure}[t]
  \centering
  \includegraphics[width=0.65\textwidth]{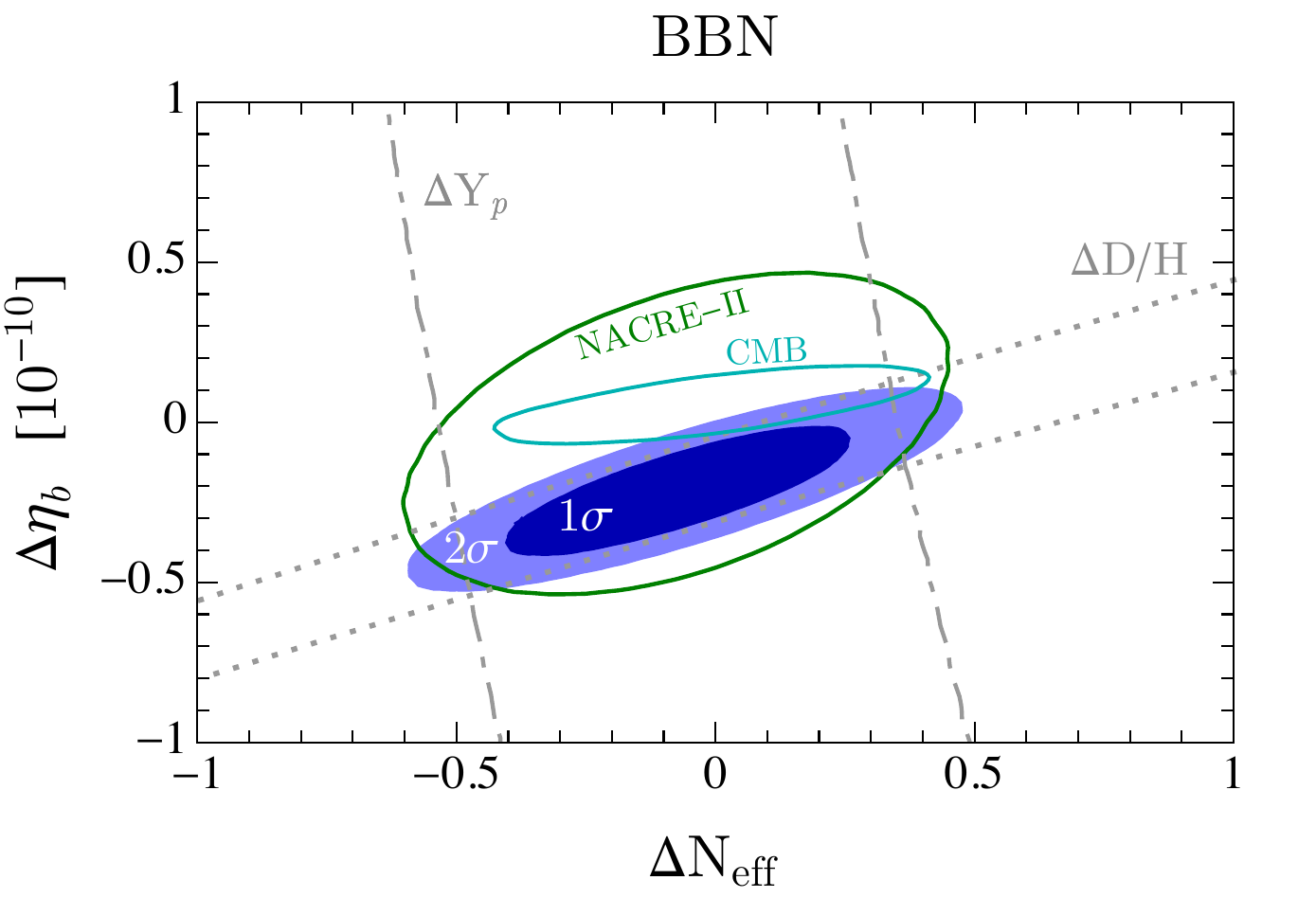}
  \caption{Regions in the $\Delta \Neff - \Delta \eta_b$ plane where the standard predictions of Big Bang nucleosynthesis are consistent with observations of the primordial helium-4 and deuterium abundances at the $1 \sigma$ (dark blue) and $2 \sigma$ (light blue) confidence level (calculated using the rates of Ref.~\cite{Coc:2015bhi}). The same analysis using the rates of Ref.~\cite{Xu:2013fha} is shown in dark green ($2 \sigma$).
Along the dot-dashed and dotted gray contours, $|\Delta Y_p| = 2$ and $|\Delta \D/\H| = 2$, respectively, using the rates of Ref.~\cite{Coc:2015bhi} (see Eq.~\ref{eq:delta_def}). 
We also compare to regions $2\sigma$-favored by recent Planck measurements of the cosmic microwave background (cyan)~\cite{Aghanim:2018eyx}. 
  \label{fig:neff_vs_eta}
   } 
\end{figure}

The above  definition of $\chi^2$ does not account for correlated uncertainties. This is justified in our baseline analysis, since we adopt the nuclear rates of Ref.~\cite{Coc:2015bhi}, in which case the correlation coefficients are much smaller than unity (see the discussion below Eq.~\ref{eq:cocetal_uncertainties}). However, this is not the case for the NACRE-II rates of Ref.~\cite{Xu:2013fha} (Eq.~\ref{eq:nacre_uncertainties}). In most regions of parameter space that we consider throughout this work, it is typically the case that $\Delta Y_p \gg \Delta \D / H$ or $\Delta Y_p \ll \Delta \D / H$, and, hence, we do not expect correlated uncertainties to significantly modify our results. To check this, we have explicitly rerun our analysis accounting for these correlations. We find that they have negligible impact in the majority of the parameter space. Hence, when displaying our results, we will ignore such correlations between $Y_p$ and $\D/\H$ even when adopting the NACRE-II rates. 

Figure~\ref{fig:neff_vs_eta} illustrates how modifications to the baryon density ($\eta_b$) and the expansion rate ($\Neff$) are constrained from measurements of the primordial densities of helium-4 and deuterium. The parameters, $\Delta \eta_b$ and $\Delta \Neff$, correspond to 
a shift of the late-time (CMB era) value of $\eta_b$ and a time-independent modification of $\Neff$ away from the SM expectation, respectively.
The evolution of $\eta_b(T)$ during nucleosynthesis for a given $\Delta\eta_b$ is evaluated using entropy conservation, 
as in Eqs.~\ref{eq:etab_evolution} and Eq.~\ref{eq:etab_initial_condition}, assuming the standard form of $\xi_\nu$.
The SM is defined by $\Neff \approx 3$, with the baryon-to-photon ratio fixed at late times to the CMB-preferred value, $\eta_b \approx 6 \times 10^{-10}$.

In our calculations of the primordial nuclei abundances, the best-fit point is defined as the value of $\Delta \Neff$ and $\Delta \eta_b$ that minimizes $\chi^2$, as defined in Eq.~\ref{eq:chi2_def}. In Fig.~\ref{fig:neff_vs_eta}, this occurs at $(\Delta \Neff, 10^{10} \Delta \eta_b) = (-0.04, -0.2)$ with a $\chi^2$ of $\chi_\text{min}^2=0.04$. In Fig.~\ref{fig:neff_vs_eta}, we show  $1 \sigma$ (dark blue) and $2 \sigma$ (light blue) regions around this best-fit point, corresponding to $\Delta \chi^2 \equiv \chi^2 - \chi_\text{min}^2 = 2.30$ and $6.18$, respectively, using the nuclear rates of Ref.~\cite{Coc:2015bhi}. 
Standard cosmology is consistent with the observed abundances within $\lesssim 2 \sigma$.  Along the dot-dashed and dotted gray contours of Fig.~\ref{fig:neff_vs_eta}, $|\Delta Y_p| = 2$ and $|\Delta \D/\H| = 2$, respectively. As discussed in Sec.~\ref{sec:dependence}, 
$Y_p$ is dominantly sensitive to $\Neff$ with only a logarithmic dependence on $\eta_b$. On the other hand, the effect of $\eta_b$ on the predicted deuterium abundance is much larger, but is degenerate with $\Neff$, since modifications to the expansion rate alter the freeze-out temperature of deuterium-burning, which can always be compensated by increasing or decreasing the burning rates with larger or smaller values of $\eta_b$. 

Compared to Ref.~\cite{Cyburt:2015mya}, our baseline analysis shows a mild preference for smaller values of $\eta_b$ and slightly smaller uncertainties 
because of our different choices for various \D-burning rates (see the discussion in Sec.~\ref{sec:procedure}). 
In particular, the rates listed in the last three rows of Table~\ref{tab:nuclear_rates} are larger and 
have smaller uncertainties compared to those used in Ref.~\cite{Cyburt:2015mya}, as reflected in Eqs.~\ref{eq:nacre_uncertainties} and~\ref{eq:cocetal_uncertainties}. In order to illustrate this point explicitly, we perform the same analysis, but instead adopt the \D-burning rates of the NACRE-II compilation~\cite{Xu:2013fha}. The corresponding $2 \sigma$-favored region is shown by the dark green contour of Fig.~\ref{fig:neff_vs_eta}. This result is similar to that of Ref.~\cite{Cyburt:2015mya}, with small discrepancies due to our different choices for the observed primordial abundances in Eq.~\ref{eq:obsyields}. We have explicitly checked that our results are consistent with those in Ref.~\cite{Cyburt:2015mya} when we adopt the nuclear rates of the NACRE-II compilation \emph{and} the inferred yields of primordial nuclei noted in Ref.~\cite{Cyburt:2015mya}. Also shown  in Fig.~\ref{fig:neff_vs_eta} are regions consistent with recent Planck measurements of the CMB~\cite{Aghanim:2018eyx}.

\section{Equilibration and Decoupling during Nucleosynthesis}
\label{sec:results}

\begin{figure}[t]
\centering
\includegraphics[width=0.49\textwidth]{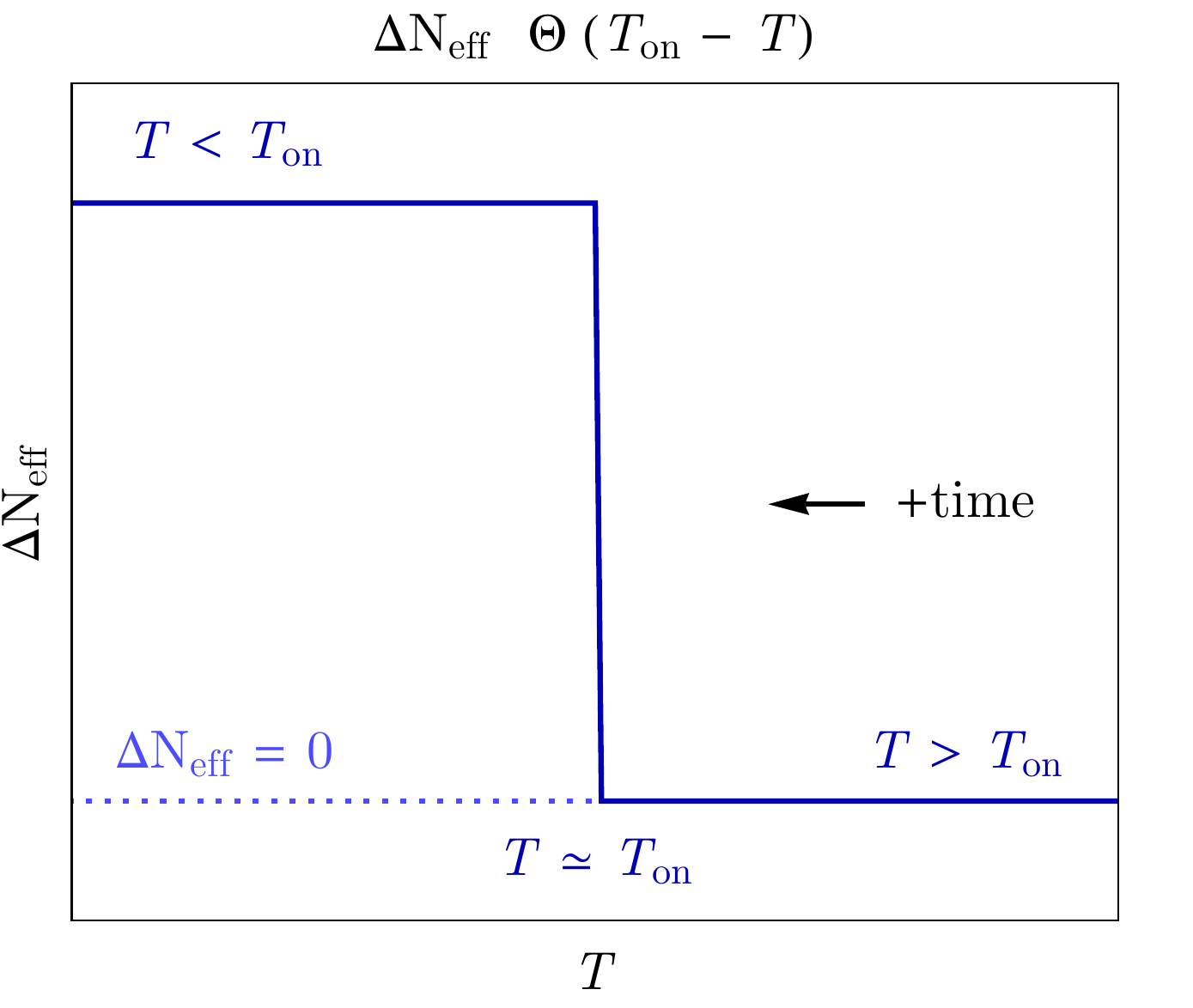}
\includegraphics[width=0.49\textwidth]{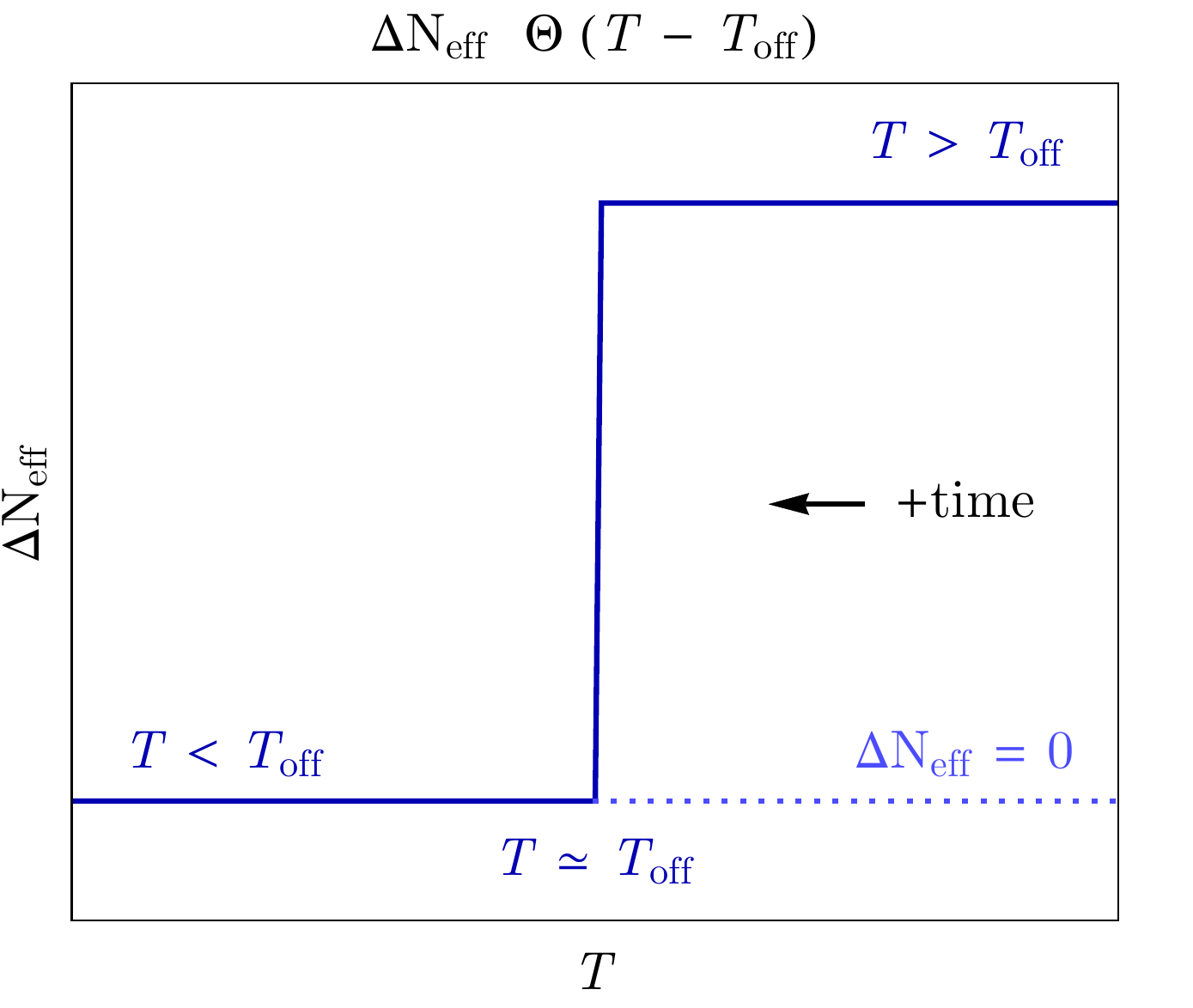}
\caption{Schematic model-independent temperature evolutions for $\Neff$ that we consider in Sec.~\ref{sec:modelindependent}. The left and right panels correspond to the left and right panels of Fig.~\ref{fig:step_bounds}. In the left panel, $\Neff$ tracks the Standard Model expectation at early times ($\DNeff = 0$). Later, a deviation to $\Neff$ ($\DNeff \neq 0$) occurs below some critical temperature, $T_\text{on}$. A similar scenario is shown in the right panel, in which a deviation turns off after the universe cools below the critical temperature, $T_\text{off}$. In Fig.~\ref{fig:pulse}, we also consider the possibility that a deviation to $\Neff$ turns on \emph{and} subsequently turns off during the epoch of primordial nucleosynthesis.
\label{fig:step_cartoon}
} 
\end{figure}

In Sec.~\ref{sec:bbn}, we reviewed how the expansion rate ($\Neff$) and the baryonic density ($\eta_b$) control the outcome of primordial nucleosynthesis
and how time-independent deviations of these quantities away from their SM expectations lead to changes in the predicted abundances of helium-4 and deuterium. In most studies of BBN, constant shifts to $\Neff$ and $\eta_b$ are constrained in this manner. 

Light and feebly-coupled dark sectors (DS) that enter equilibrium with the SM bath below the temperature of neutrino-photon decoupling naturally lead to deviations in $\Neff$ and $\eta_b$ that are effectively temperature- or time-dependent, as shown in Figs.~\ref{fig:schematic_neff_evol}-\ref{fig:neff_and_etab_evol_gamma}. In this case, modifications to primordial nucleosynthesis may occur in specific time/temperature intervals, and adapting bounds from previous studies is not straightforward. In this section, we discuss this more general scenario, in which $\Neff$ and $\eta_b$ evolve non-trivially as the universe adiabatically cools below critical temperatures and mass-thresholds of the DS. In Sec.~\ref{sec:modelindependent}, we present constraints from considerations of BBN in a model-independent manner, for specific simplified forms of $\Delta \Neff(T)$ (where $T$ is the temperature of the photon bath) and fixing $\Delta \eta_b = 0$. In Sec.~\ref{sec:modeldependent}, we consider a few concrete models that predict non-standard temperature evolution of $\Delta \Neff (T)$ \emph{and} $\Delta \eta_b(T)$ and discuss how the predicted nuclear abundances are modified in relevant regions of parameter space.

\begin{figure}[t]
  \centering
  \includegraphics[width=0.49\textwidth]{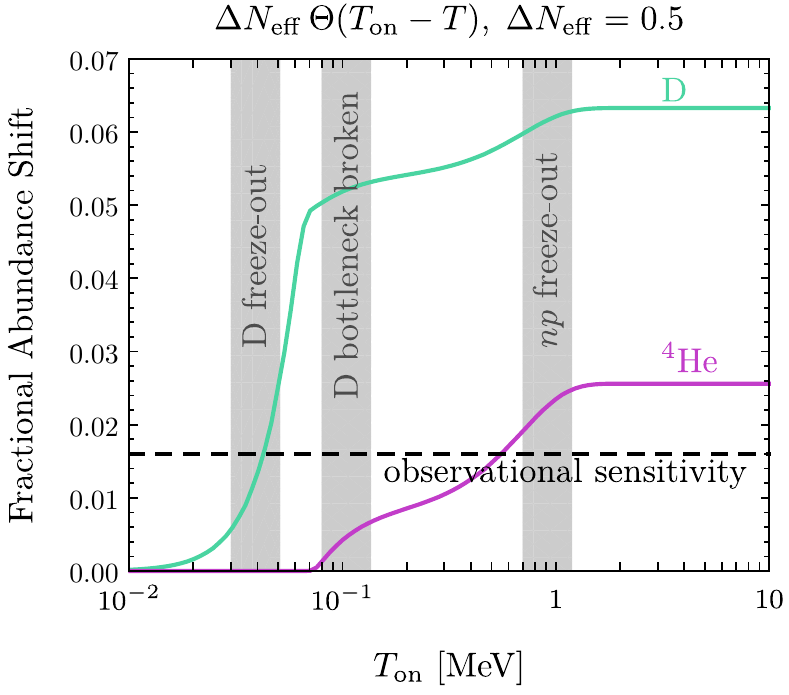}
  \includegraphics[width=0.49\textwidth]{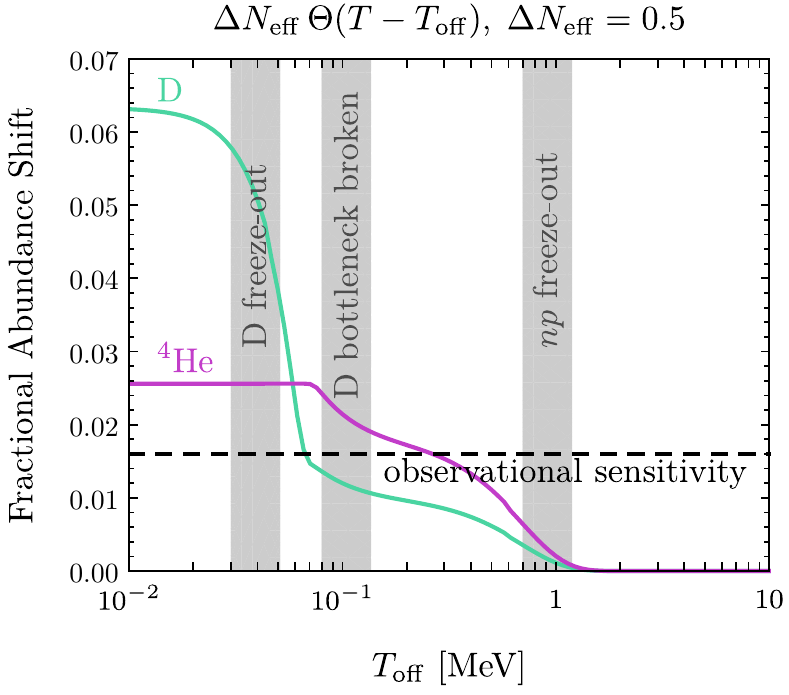}
  \caption{Impact of step-like modifications to $\Neff$ on the fractional yields of \he4 and \D as a function of the 
      transition temperature ($T_{\mathrm{on}}$ or $T_{\mathrm{off}}$) for fixed amplitude, $\DNeff = 0.5$.
      The left (right) panel corresponds to the time-evolution shown in the left (right) panel of Fig.~\ref{fig:step_cartoon}.
      The fractional variations of \he4 and \D yields (compared to $\DNeff =  0$) are shown as purple and green lines, respectively.
      Key epochs in primordial nucleosynthesis (neutron-proton freeze-out, the deuterium bottleneck and freeze-out) are highlighted in gray. The horizontal black dashed line indicates 
      the approximate sensitivity of current observations. 
  \label{fig:yield_variation_step_like}
} 
\end{figure}
\begin{figure}[t]
  \centering
   \includegraphics[width=0.49\textwidth]{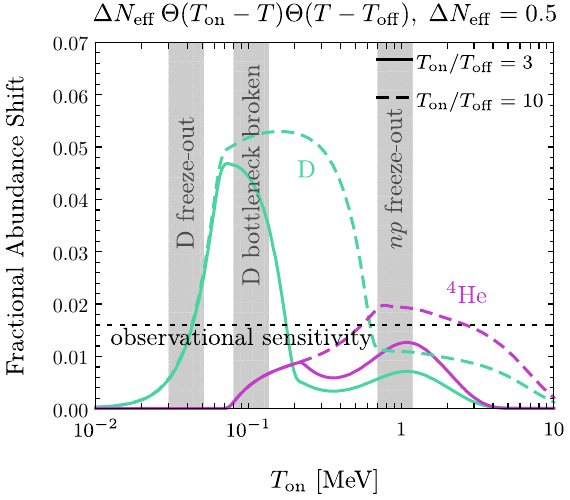}
  \caption{
As in Fig.~\ref{fig:yield_variation_step_like}, but for a pulse-like modification to $\Neff$ (see Eq.~\ref{eq:Tpulse}). Its impact on the yields of \he4 and \D as a function of the 
    transition temperature, $T_{\mathrm{on}}$, is shown for $T_{\mathrm{on}}/T_{\mathrm{off}} = 3$ and $10$ (solid and dashed colored lines, respectively) 
    and for a fixed pulse amplitude of $\DNeff = 0.5$.   \label{fig:yield_variation_pulse_like}
} 
\end{figure}

\subsection{Model-Independent Results}
\label{sec:modelindependent}

In this section, we only focus on model-independent modifications to the expansion rate and adopt a few representative ``temperature waveforms'' for the functional form of $\Neff(T)$. The first possibility that we consider is that $\Neff$ tracks the SM expectation of $\Neff \approx 3$ until some later temperature, $T_\text{on}$, at which point a deviation turns on, i.e., $\Delta \Neff \neq 0$ for $T \lesssim T_\text{on}$. This is shown in the left panel of Fig.~\ref{fig:step_cartoon}.  As discussed in Sec.~\ref{sec:neutrinocoupling}, this scenario is motivated by a sub-MeV DS that recouples and later decouples with the neutrino bath after neutrino-photon decoupling. 

We parametrize this temperature-behavior as
\beq
\label{eq:Ton}
\Delta \Neff (T) = \Delta \Neff ~ \Theta (T_\text{on} - T) ~~ \text{(step-like)}
~,
\eeq
where the temperature-dependence is encapsulated in the Heaviside step-function, $\Theta$. We also consider the possibility that $\Delta \Neff \neq 0$ at the beginning of nucleosynthesis, but that this deviation turns off at some critical temperature, $T_\text{off}$, i.e.,
\beq
\label{eq:Toff}
\Delta \Neff (T) = \Delta \Neff ~ \Theta (T - T_\text{off}) ~~ \text{(step-like)}
~,
\eeq
as shown in the right panel of Fig.~\ref{fig:step_cartoon}. In general, dark sectors that recouple and later decouple with the SM bath during nucleosynthesis can result in more intricate behaviors of $\Delta \Neff (T)$. We therefore also investigate the generalization of Eqs.~\ref{eq:Ton} and \ref{eq:Toff}, 
\beq
\label{eq:Tpulse}
\Delta \Neff (T) = \Delta \Neff ~ \Theta (T_\text{on} - T) ~ \Theta (T - T_\text{off}) ~~ \text{(pulse-like)}
~.
\eeq
This corresponds to a sudden pulse-like modification to $\Neff$ within the temperature range of $T_\text{off} \lesssim T \lesssim T_\text{on}$. 

If these deviations to $\Neff$ occur in the temperature range $\order{10} \text{ keV} \lesssim T_\text{on,off} \lesssim \order{1} \text{ MeV}$, then important epochs of nucleosynthesis are potentially modified. Figures~\ref{fig:yield_variation_step_like} and \ref{fig:yield_variation_pulse_like} show the fractional shift (compared to $\DNeff = 0$) to the helium and deuterium abundances for these various temperature evolutions of $\DNeff (T)$, fixing the amplitude to $\DNeff = 0.5$. The final abundance of $\he4$ is modified in the standard way (see Sec.~\ref{sec:dependence}) if $\DNeff \neq 0$ during neutron-proton freeze-out ($100 \text{ keV} \lesssim T \lesssim \text{ MeV}$), while the deuterium abundance is most sensitive to the expansion rate near the freeze-out of \D-burning process ($50 \text{ keV} \lesssim T \lesssim 100 \text{ keV}$). However, it is important to note that drastic modifications to the neutron-to-proton ratio do also affect deuterium yields since its final abundance ultimately depends upon the neutrons that do not end up bound into helium nuclei. This effect can be seen by the feature in the green contours near $T \sim \MeV$ in Figs.~\ref{fig:yield_variation_step_like} and \ref{fig:yield_variation_pulse_like}, which shows that the deuterium abundance is indeed slightly affected by modifications to the expansion rate that occur well before the deuterium bottleneck is overcome.

\begin{figure}[t]
\centering
\includegraphics[width=0.49\textwidth]{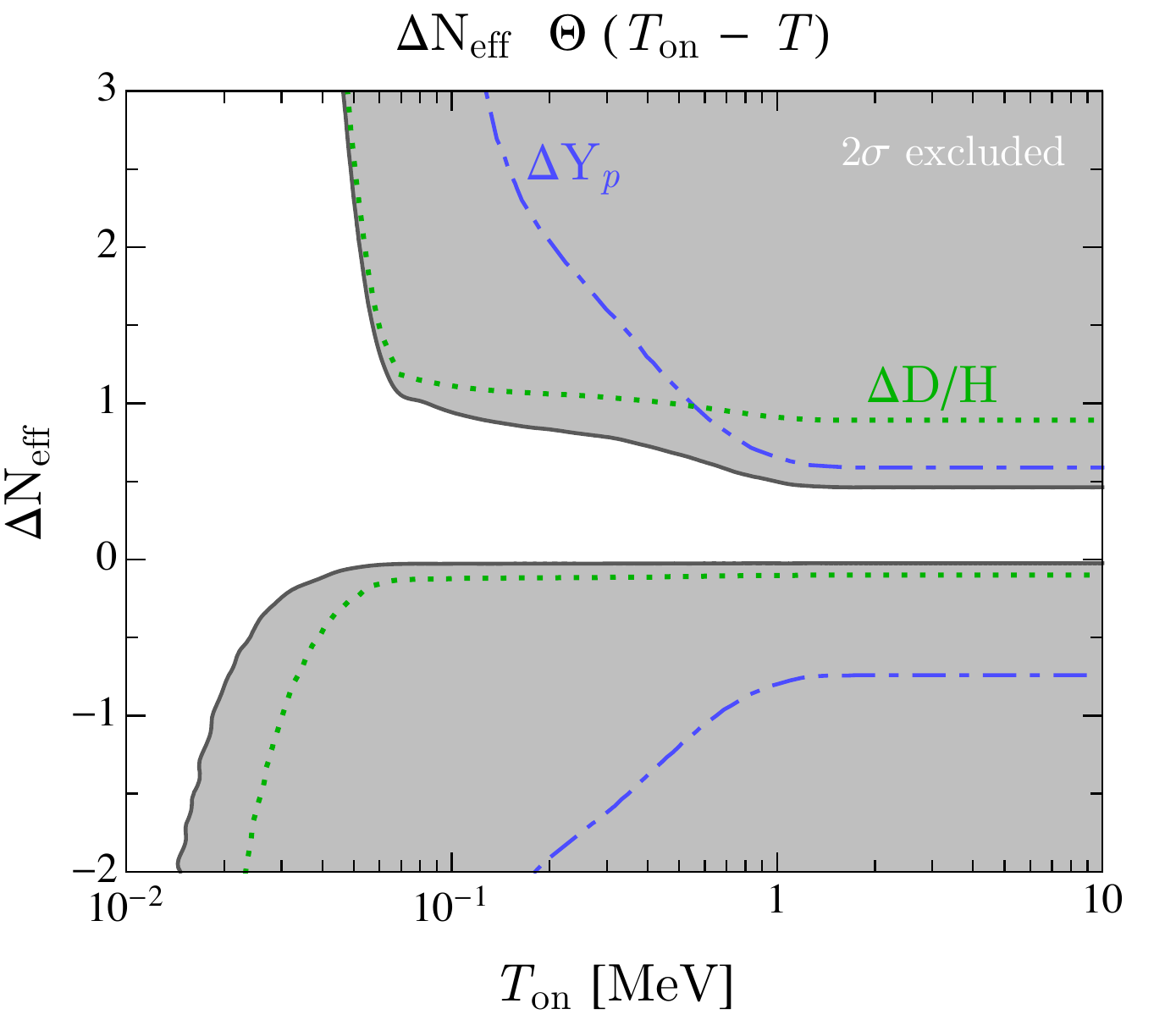}
\includegraphics[width=0.49\textwidth]{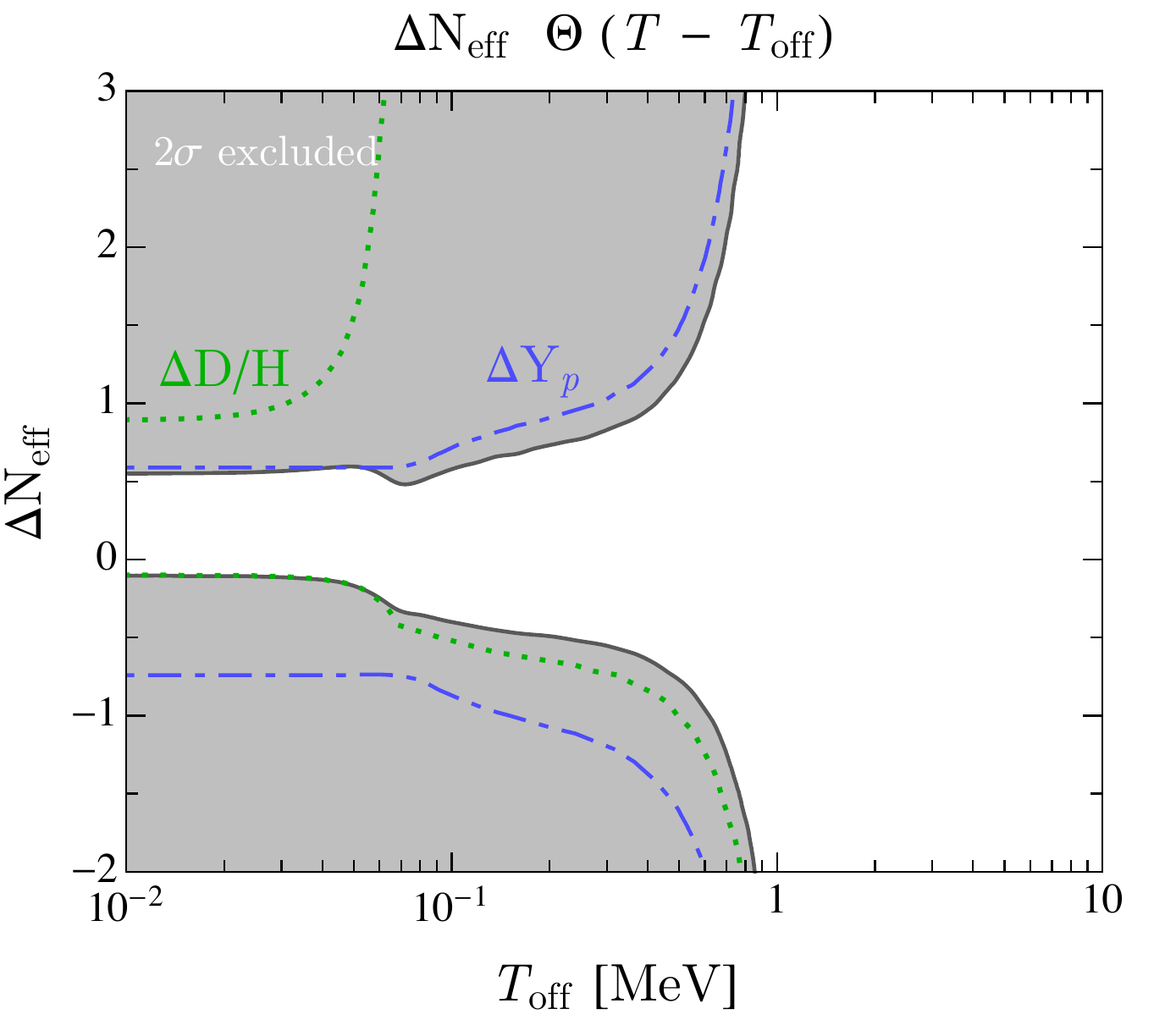}
\caption{Model-independent constraints on the step-function temperature evolution of $\Neff$ shown in Fig.~\ref{fig:step_cartoon}. Shown in gray are regions of parameter space that are inconsistent (within $2 \sigma$) with observations of the primordial helium-4 and deuterium abundances (calculated using the rates of Ref.~\cite{Coc:2015bhi}). Along the blue dot-dashed and green dotted  contours, $|\Delta Y_p| = 3$ and $|\Delta \D/\H| = 3$, respectively (see Eq.~\ref{eq:delta_def}). 
\label{fig:step_bounds}
} 
\end{figure}

In the left and right panels of Fig.~\ref{fig:step_bounds}, we highlight regions of parameter space that are inconsistent with considerations of BBN in the $\Delta \Neff-T_\text{on}$ and $\Delta \Neff-T_\text{off}$ plane, respectively, assuming a step-like temperature evolution of $\DNeff (T)$, as parametrized by Eqs.~\ref{eq:Ton} and \ref{eq:Toff}.  Also shown as blue and green contours are parameters for which $|\Delta Y_p | = 3$ and $|\Delta \D / \H| = 3$, respectively (see Eq.~\ref{eq:delta_def}). For this analysis, we adopt the same prescription as described in Sec.~\ref{sec:standard} and exclude parameters that predict $\Delta \chi^2 \gtrsim 6.18$. Throughout, we have fixed the baryon density, $\eta_b$, such that it agrees with the measured value at the time of recombination, as given by Eq.~\ref{eq:etab_initial_condition}, and tracks the standard evolution at earlier times. If the cosmological expansion rate is modified during $50 \text{ keV} \lesssim T \lesssim \text{MeV}$, then our calculated bound on $\Neff$ is similar to the standard constraint presented in Fig.~\ref{fig:neff_vs_eta}, i.e., $|\Delta \Neff | \lesssim 0.5$. However, if modifications to the expansion rate occur when $T \lesssim 50 \text{ keV}$ or $T \gtrsim \text{MeV}$, then deviations as large as $|\Delta \Neff | \gtrsim \order{1}$ are consistent with the measured abundances of helium-4 and deuterium.

\begin{figure}[t]
\centering
\includegraphics[width=0.47\textwidth]{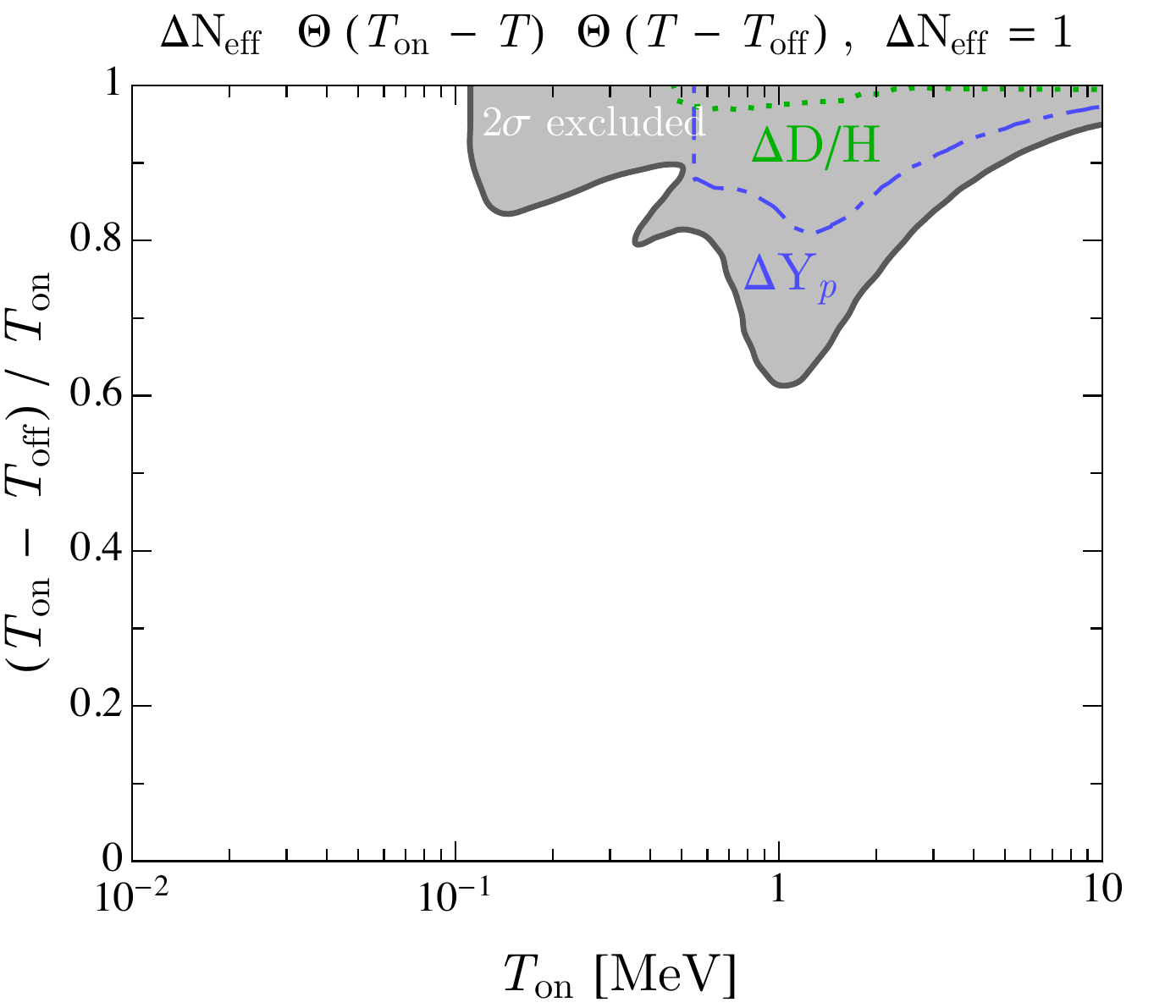}
\includegraphics[width=0.47\textwidth]{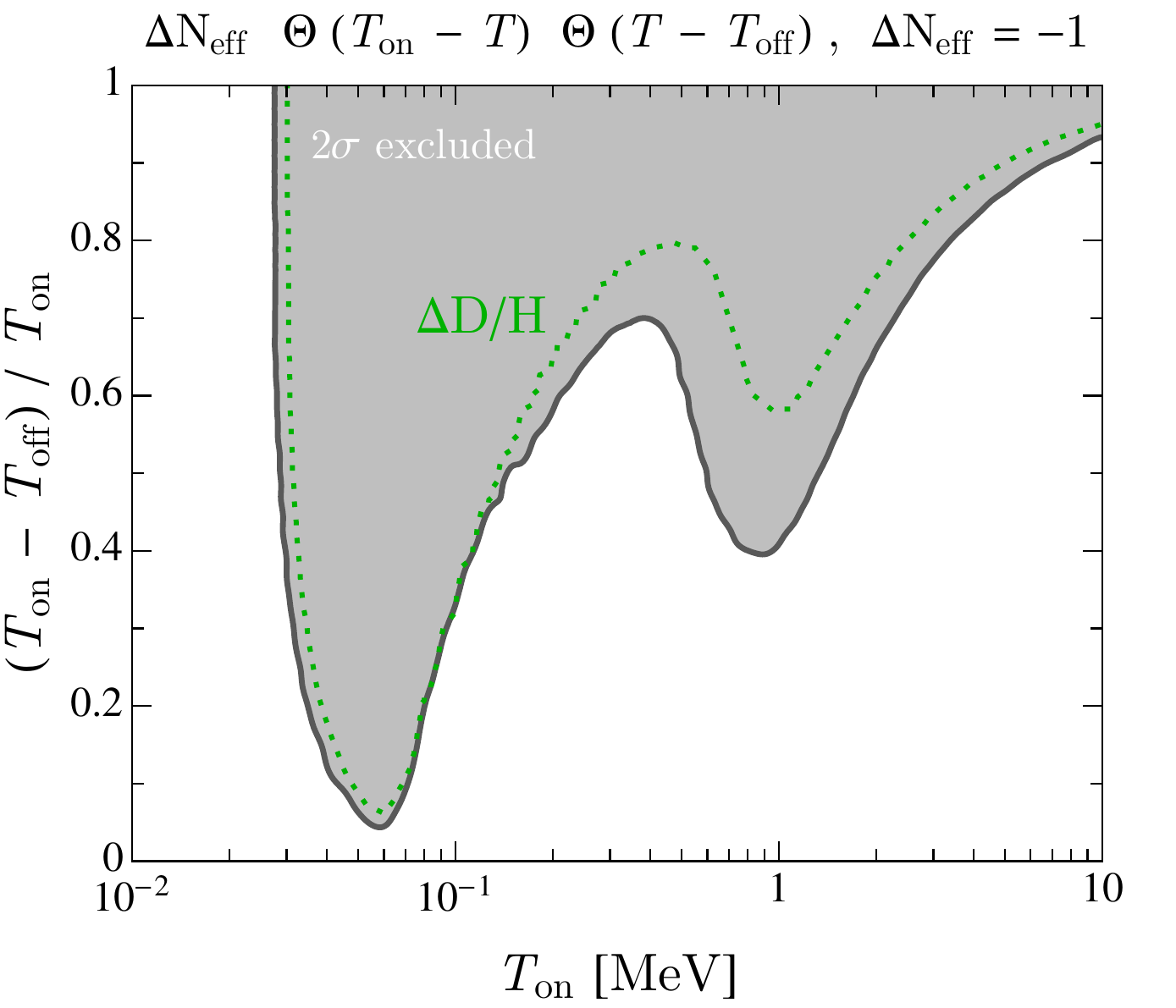}
\includegraphics[width=0.47\textwidth]{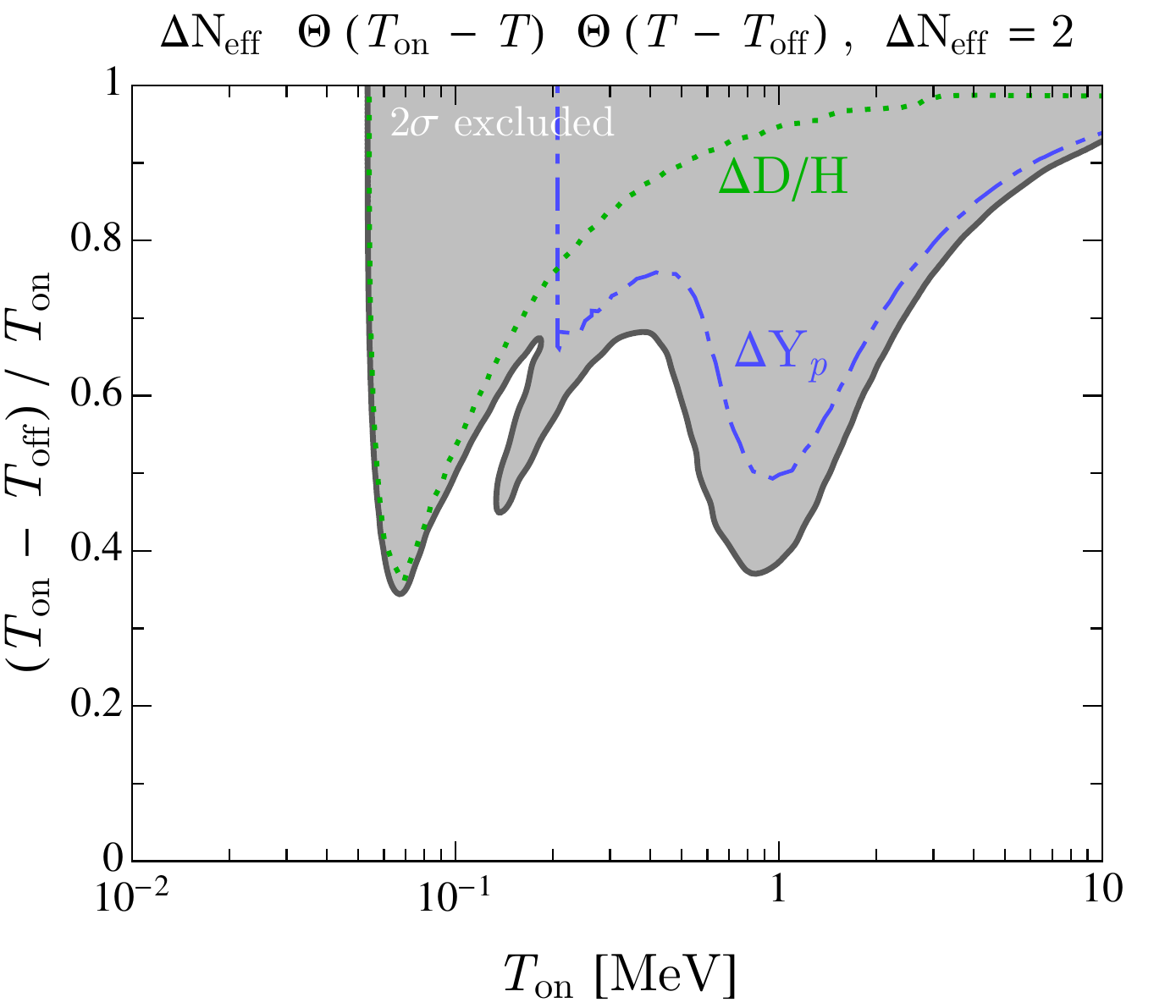}
\includegraphics[width=0.47\textwidth]{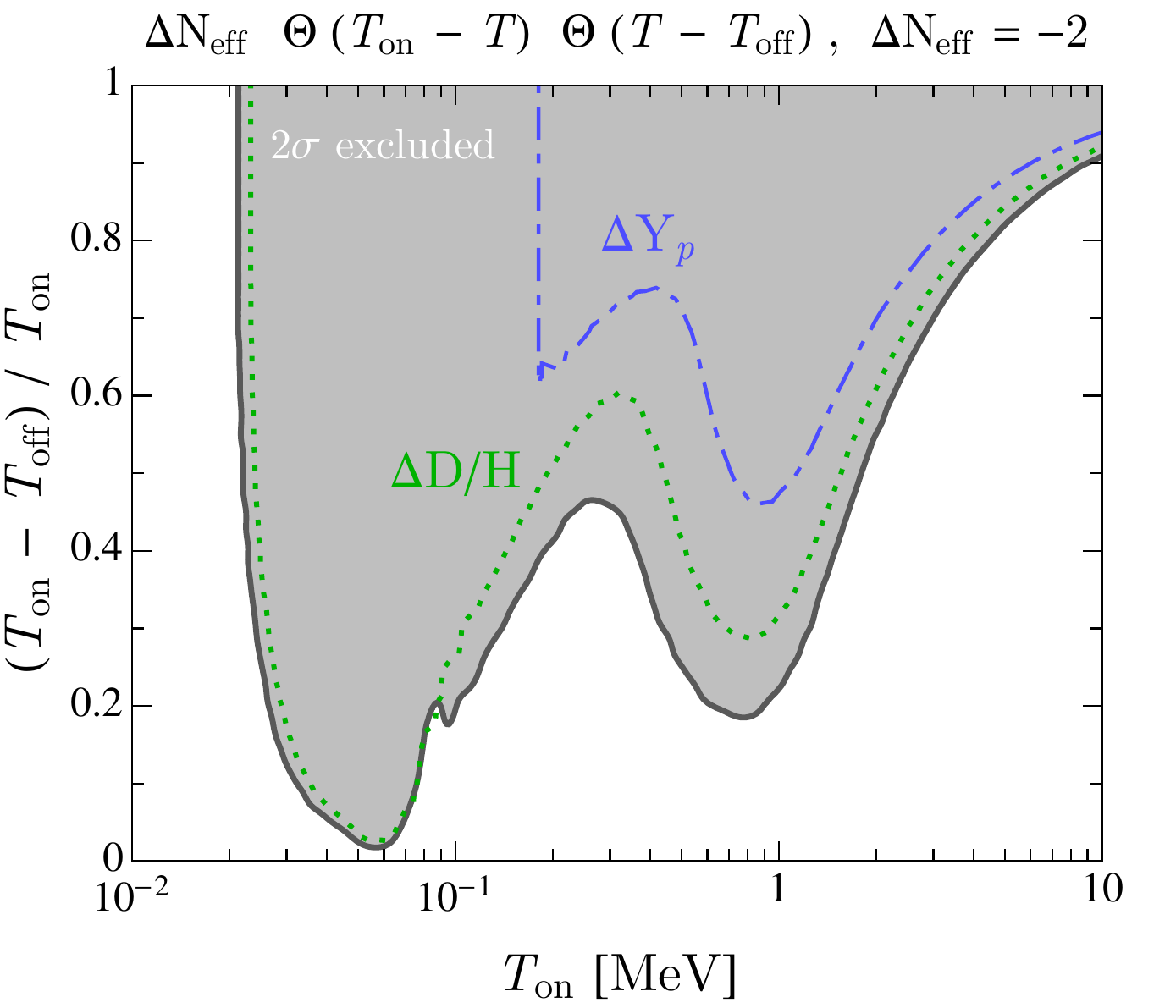}
\caption{Model-independent constraints on the pulse-like temperature evolution of $\Neff$ (see Eq.~\ref{eq:Tpulse}). Shown in gray are regions of parameter space that are inconsistent (within $2 \sigma$) with observations of the primordial helium-4 and deuterium abundances (calculated using the rates of Ref.~\cite{Coc:2015bhi}). Along the blue dot-dashed and green dotted  contours, $|\Delta Y_p| = 3$ and $|\Delta \D/\H| = 3$, respectively (see Eq.~\ref{eq:delta_def}). Significant deviations to the expansion rate are allowed to occur during nucleosynthesis, provided that this occurs after neutron-proton freeze-out and before deuterium burning ($100 \text{ keV} \lesssim T \lesssim \MeV$). See Sec.~\ref{sec:modelindependent} for a more detailed discussion of the various temperature-dependent features of these exclusions.
\label{fig:pulse}
} 
\end{figure}

As discussed in Sec.~\ref{sec:dependence}, measurements of primordial deuterium are largely sensitive to $\Neff$  deviations that occur slightly before the end of nucleosynthesis. This is illustrated in the left panel of Fig.~\ref{fig:step_bounds}, which shows that scenarios in which $\Delta \Neff \neq 0$ only for $T \lesssim \text{few} \times \order{100} \text{ keV}$ are predominantly constrained by measurements of the deuterium abundance. Also apparent in Fig.~\ref{fig:step_bounds} is the asymmetrical importance of $Y_p$ and $\D / \H$ for constraining $\DNeff >0$ or $\DNeff < 0$. This can be understood from the fact that the SM ($\Delta \Neff = \Delta \eta_b = 0$) predicts a slight overabundance in helium and a more significant underabundance in deuterium (see Eq.~\ref{eq:SMDelta}). As a result, most of the constraining power for $\Delta \Neff < 0$ comes from measurements of $\D / \H$, since such cosmologies lead to an even more significant underabundance of deuterium, as explained in Sec.~\ref{sec:dependence}.

Constraints on pulse-like evolutions of $\DNeff$ are shown in Fig.~\ref{fig:pulse} (see Eq.~\ref{eq:Tpulse}). In each panel of Fig.~\ref{fig:pulse}, we vary the fractional width of the $\Neff$ pulse, $(T_\text{on} - T_\text{off})/T_\text{on}$, as well as the temperature at which the pulse turns on, $T_\text{on}$, for different choices of the pulse height, $\DNeff = \pm 1$, $\pm 2$. Regions of parameter space that are inconsistent with the predictions of BBN are shown in solid gray. This analysis is nearly identical to ones shown earlier in this work, but since we are now varying parameters in the three-dimensional parameter space spanned by $T_\text{on}$, $T_\text{off}$, and $\Neff$, we exclude parameters at the $2 \sigma$ level if $\Delta \chi^2 \gtrsim 8.03$. Figure~\ref{fig:pulse} illustrates that significant modifications to the expansion rate are allowed to occur during nucleosynthesis, provided that these deviations happen after neutron-proton freeze-out and before deuterium burning, i.e., when the temperature of the photon bath is $100 \text{ keV} \lesssim T \lesssim \MeV$. As in Fig.~\ref{fig:step_bounds}, measurements of the deuterium abundance are typically more constraining for $\DNeff < 0$. On the other hand, for $\DNeff > 0$, the relative importance of helium and deuterium strongly depends on when the deviation to $\Neff$ takes place, which is most clearly illustrated in the bottom-left panel of Fig.~\ref{fig:pulse}. As discussed above and shown in Fig.~\ref{fig:yield_variation_pulse_like}, deuterium yields are dominantly sensitive to the cosmological expansion rate near the end of nucleosynthesis, with a subdominant sensitivity to earlier epochs near neutron-proton freeze-out. The structure of the exclusion at lower values of $(T_\text{on} - T_\text{off})/T_\text{on}$ 
(corresponding to narrower pulses) visible in the left column of Fig.~\ref{fig:pulse} can be understood by comparison 
with Fig.~\ref{fig:yield_variation_pulse_like}.

\subsection{Model-Specific Results}
\label{sec:modeldependent}
In this section, we consider two concrete minimal models in which a DS equilibrates with the neutrino or photon bath after neutrino-photon decoupling via decays and inverse decays of a light bosonic mediator, $\p$: 
$\p \leftrightarrow \nu\nu \; \mathrm{or}\; \p \leftrightarrow \gamma\gamma$. These processes can be realized by 
the interactions given in Eqs.~\ref{eq:nu_interaction} and~\ref{eq:gamma_interaction}
for decays to SM neutrinos and photons, respectively. Throughout this section, we use the notation introduced in Sec.~\ref{sec:dark_sector_equilibration}. 
Since equilibration and decoupling between the DS and the SM do not occur instantaneously, we solve the Boltzmann equations to find $\Neff (T)$ and $\eta_b (T)$ (see Eqs.~\ref{eq:energy_density_boltzmann1} and \ref{eq:simple_collision_term}).
The relevant collision terms are given in Appendix~\ref{sec:collision_terms}. Solutions of the Boltzmann equations for different illustrative values of 
the common DS mass scale, $\mds$, and DS degrees of freedom, $g_*^\ds$, are shown in 
Figs.~\ref{fig:neff_evol_nu} and \ref{fig:neff_and_etab_evol_gamma} for neutrino and photon couplings, respectively. 

\begin{figure}[t]
  \centering
\includegraphics[width=10cm]{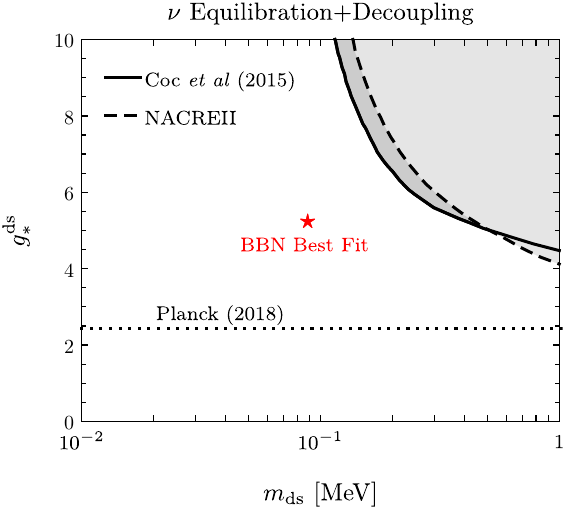}
\caption{Nucleosynthesis constraints on the equilibration of a cold dark sector ($\xi^0_\ds = 0.3$) with neutrinos (assuming that this occurs after
  neutrino-photon decoupling), as a function of the dark sector degrees of freedom ($g^\ds_*$) and the dark sector mass scale ($m_\ds$). The $2 \sigma$-excluded region (shaded gray) is shown for two different parametrizations of \D-burning rates: the NACREII compilation (dashed line)~\cite{Xu:2013fha} and those from \cocetal~\cite{Coc:2015bhi} (solid line). 
The horizontal dotted line shows the Planck bound on $\Neff$ (without fixing $Y_p$ to the standard BBN value), which excludes the region above the line. 
The best-fit model (for considerations of BBN alone) is indicated by the red star.
  \label{fig:neff_bounds_neutrino_equilibration}}
\end{figure}

We evaluate the light element abundances for these various cosmologies 
and compare them to the observed values, as in Sec.~\ref{sec:bbn}. 
In Fig.~\ref{fig:neff_bounds_neutrino_equilibration}, we show regions of parameter space for the neutrino-coupled model 
that are excluded from considerations of BBN, as a function of the 
DS mass scale ($m_\ds$) and the DS effective number of relativistic degrees of freedom ($g_*^\ds$) for an initial temperature ratio of $\xi^0_\ds = 0.3$.
The $ 2\sigma$-excluded regions (solid gray) correspond to $\Delta \chi^2 = 6.18$. 
Gray regions outlined 
by solid and dashed lines are obtained using 
two different sets of \D-burning rates, as discussed in Sec.~\ref{sec:procedure}.
The equilibration and decoupling of a cold DS with the neutrino bath gives rise to a 
step-like deformation in $\Neff$ (see Eq.~\ref{eq:Ton}) occurring at $T_{\mathrm{on}} \sim m_{\ds}/5$, as 
shown in Fig.~\ref{fig:neff_evol_nu}. Hence, the shape of these exclusions can be understood from the left 
panel of Fig.~\ref{fig:yield_variation_step_like}. For larger DS masses ($m_\ds \gg \text{MeV}$), $T_\text{on} \gtrsim \text{MeV}$ and $\Neff$ 
is modified during every key epoch of nucleosynthesis, which is very strongly constrained for $g_*^\ds \gtrsim \text{few}$.
As the DS mass scale is decreased, the expansion history is altered for a correspondingly shorter 
time/temperature interval during nucleosynthesis. Hence, DS masses lighter than $\sim 100 \text{ keV}$ are typically 
unconstrained from considerations of BBN, since in this case $\Neff$ is modified only after the most important processes of nucleosynthesis have concluded.
Precisely the same behavior is seen in the model-independent constraint derived in the previous 
section and shown in the left panel of Fig.~\ref{fig:step_bounds}. In fact, the model-independent result 
can be used to approximate the bound on $g_*^\ds$ in Fig.~\ref{fig:neff_bounds_neutrino_equilibration} 
by using $T_{\mathrm{on}} \sim m_{\ds}/5$ and Eq.~\ref{eq:neff_after_decoupling_nu_equil} to translate $\Delta \Neff$ into $g_*^\ds$.

The horizontal dotted line in Fig.~\ref{fig:neff_bounds_neutrino_equilibration} 
shows the $2 \sigma$ upper bound on $g_*^\ds$ derived from Eq.~\ref{eq:neff_after_decoupling_nu_equil} and 
Planck measurements of $\Neff$ at recombination\footnote{We use the Planck result from the joint fit for $\Neff$ and $Y_p$, which does not assume standard BBN. 
In this case, marginalizing over $Y_p$ in the resulting likelihood gives $\Neff = 2.99^{+0.43}_{-0.40}$ at $2 \sigma$
for the \texttt{Planck TT,TE,EE+lowE+lensing+BAO} datasets~\cite{Aghanim:2018eyx}.}~\cite{Aghanim:2018eyx}:
\beq
g_*^\ds \lesssim 2.4 ~~ (\text{CMB, DS-$\nu$ equilibration}) 
~.
\eeq
For this neutrino-coupled scenario, the CMB bound is always more stringent than 
the one derived from considerations of BBN. There are several reasons for this. 
First, the CMB is sensitive to the expansion rate at much later times, when the presence of the DS has maximally modified $\Neff$. 
In contrast, bounds derived from primordial nucleosynthesis depend on precisely when modifications to $\Neff$ commence. 
For instance, if the mass of a cold neutrino-coupled DS particle is smaller than the temperature at which \D-burning processes freeze out, $\Neff$ is SM-like throughout the most important epochs of nucleosynthesis. Hence, bounds derived from BBN are weakened for $\mds \lesssim 100 \text{ keV}$.
Second, these constraints also depend on the particular set of \D-burning rates adopted in the analysis, as discussed in Sec.~\ref{sec:bbn}.
When nuclear rates from the NACRE-II compilation~\cite{Xu:2013fha} are adopted, theoretical uncertainties for deuterium yields are large and matching to the observed nuclei abundances (Eq.~\ref{eq:obsyields}) leads to a preference for a SM-like expansion rate ($\Neff \approx 3$).
Instead, when rates from \cocetal~\cite{Coc:2015bhi} are utilized, theoretical uncertainties of deuterium yields are significantly reduced (compare Eqs.~\ref{eq:nacre_uncertainties} and~\ref{eq:cocetal_uncertainties}) and the larger \D-burning rates lead to a corresponding preference for a larger expansion rate (compared to NACRE-II). Both of these effects tend to substantially weaken the BBN constraints on $g_*^\ds$, compared to those derived from the CMB. 
The point in parameter space that is preferred from considerations of BBN alone
(for the \cocetal rates~\cite{Coc:2015bhi}), shown as the red star in Fig.~\ref{fig:neff_bounds_neutrino_equilibration}, realizes a 
larger expansion and provides a better fit to the observed \D abundance than the SM if $\eta_b$ is fixed to the CMB-preferred value. 
This parameter point, however, is robustly ruled out by Planck~\cite{Aghanim:2018eyx}, since it corresponds to $\DNeff \sim 0.8$ at recombination.

\begin{figure}[t]
  \centering
\includegraphics[width=10cm]{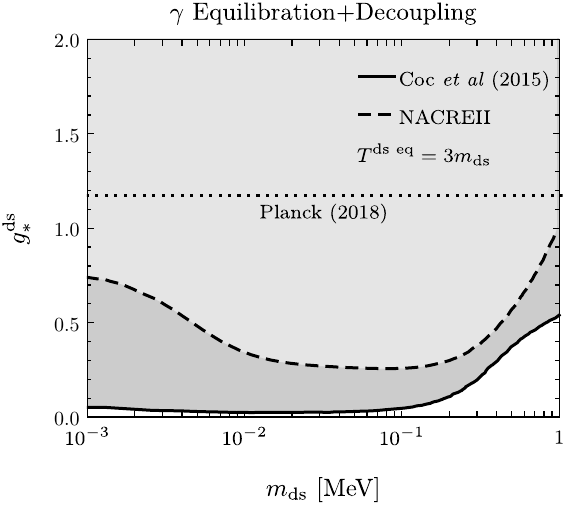}
\caption{
Nucleosynthesis constraints on the equilibration of a cold dark sector ($\xi^0_\ds = 0.3$) with the photon bath (assuming that this occurs after
  neutrino-photon decoupling), as a function of the dark sector degrees of freedom ($g^\ds_*$) and the dark sector mass scale ($m_\ds$). We have fixed the equilibration temperature to be $\TXeq \approx 3 m_{\ds}$, such that 
  these dark sector degrees of freedom are relativistic at this time.
  The $2 \sigma$-excluded region is shown in gray. The baryon density, $\eta_b$, is determined 
  by requiring consistency with the late-time CMB measurement (see Eq.~\ref{eq:etab_evolution}). 
  The resulting constraints are shown for two parametrizations of \D-burning rates: those from the NACRE-II~\cite{Xu:2013fha} compilation (dashed line) and \cocetal~\cite{Coc:2015bhi} (solid line).
  \label{fig:neff_bounds_photon_equilibration}}
\end{figure}

Compared to neutrino-coupled models, the relative importance of the CMB and BBN in constraining light dark sectors is reversed in the photon-equilibration case. 
This is shown in Fig.~\ref{fig:neff_bounds_photon_equilibration}. Here, we have fixed 
the DS-photon equilibration temperature such that $T^{\text{ds eq}} = 3 \, m_\ds$, which ensures that the DS degrees of freedom 
are relativistic at equilibration and therefore contribute fully to the 
expansion rate. This choice of parameters allows for the 
  largest viable range of DS masses to be considered, as we justify below. 
  The solid and dashed lines indicate BBN constraints that are obtained by using the nuclear rates from \cocetal~\cite{Coc:2015bhi} and 
NACRE-II~\cite{Xu:2013fha}, respectively. 
The dotted line in Fig.~\ref{fig:neff_bounds_photon_equilibration} shows the $2 \sigma$ upper bound on $g_*^\ds$ that is derived from Eq.~\ref{eq:neff_after_decoupling_photon_equil} and Planck measurements of $\Neff$ at recombination~\cite{Aghanim:2018eyx}:
\beq
g_*^\ds \lesssim 1.2 ~~ (\text{CMB, DS-$\gamma$ equilibration})
~. 
\eeq

As shown in Sec.~\ref{sec:photoncoupling} and Fig.~\ref{fig:neff_and_etab_evol_gamma}, modifications 
to the expansion rate at late times (relevant for CMB measurements) can be much smaller than 
those during nucleosynthesis in these models. As a result, the measured abundances of \he4 and \D provide the 
leading constraints on such new degrees of freedom. In addition to its effect on the expansion rate, 
 equilibration and decoupling of a DS with the photon bath modifies the evolution of the baryon-to-photon ratio, $\eta_b$.
If we were to \emph{ignore} this effect on $\eta_b$, then BBN would only constrain 
  a finite interval of $\mds$ (for a fixed value of $T^{\text{ds eq}}/\mds$), reflecting the pulse-like modification of the expansion 
rate, as shown in Fig.~\ref{fig:neff_and_etab_evol_gamma}.
If this pulse in the expansion rate occurs well before or well after nucleosynthesis (corresponding to large and small $\mds$, respectively), then the predicted yields are not modified relative to the SM case. 
However, 
once we account for modifications to $\eta_b$, the resulting bounds extend to much lower DS masses.
The baryon-to-photon ratio is independently measured at recombination, and hence we fix $\eta_b$ at earlier times 
as in Eq.~\ref{eq:etab_evolution}. Dark sector-photon equilibration and decoupling results in an irreducible increase to the comoving entropy. As a result, before the onset of nucleosynthesis, $\eta_b$ has to be \emph{larger} than in a standard cosmology, if it is fixed to the CMB-preferred value at later times.
Hence, even if DS-photon equilibration and decoupling occur well after nucleosynthesis has concluded, the primordial abundances of nuclei are still modified 
by the non-standard value of $\eta_b$ during nucleosynthesis.
As a result, considerations of BBN lead to constraints that extend down to sub-keV DS masses. 

The strength of the BBN constraint on $g_*^\ds$ in Fig.~\ref{fig:neff_bounds_photon_equilibration} can be understood by comparing the 
maximum deviation of $\Neff$ during photon equilibration (Eq.~\ref{eq:Neff2}) to the  
constraint on standard \emph{constant} shifts of $\DNeff \lesssim 0.5$ in Fig.~\ref{fig:neff_vs_eta}: 
  solving for $g_*^\ds$, one finds that the expected bound should lie near $g_*^\ds \sim 0.1$, 
which is borne out in the complete calculation of Fig.~\ref{fig:neff_bounds_photon_equilibration}. 
The observed abundances of light nuclei constrain $g_*^\ds < 1$, thereby excluding even a minimal 
DS that contains a single scalar degree of freedom that equilibrates with photons while relativistic and after neutrino-photon decoupling. 
Instead, if equilibration occurs while the DS is semi- or non-relativistic ($\TXeq < 3\mds$), 
then the \emph{effective} value of $g_*^\ds$ can be smaller than unity even if there exist several light states in the DS. 
This process of semi-relativistic equilibration leads to a smaller shift in $\Neff$ 
than what is expected from the estimates in Sec.~\ref{sec:dark_sector_cosmo}~\cite{Berlin:2018ztp}. However, this scenario also requires a coincidence 
of scales, i.e., $\mds/4 \lesssim \TXeq < 3 \mds$. This is because for $T \lesssim \mds/4$, the DS-SM equilibration rate 
no longer increases relative to the Hubble parameter with the expansion of the universe, while for $T \gtrsim 3\mds$, $g_*^\ds \geq 1$ and equilibration is relativistic. Furthermore, even for scenarios in which $\TXeq \sim \mds/4$ (corresponding to $g_*^\ds \gtrsim 0.25$ during equilibration), there remains a slight tension with the BBN bounds of Fig.~\ref{fig:neff_bounds_photon_equilibration}.

Up to this point, we have ignored the presence of the photon plasma frequency.
 For temperatures above the electron mass ($T \gtrsim m_e$), 
the effective mass of transverse photon excitations is $m_t^2 \approx 4\pi \alpha T^2/9$, which 
falls quickly to $m_t^2 \approx 4\pi \alpha n_e/m_e$ as $T$ drops below $m_e$~\cite{Braaten:1993jw}. In 
certain regions of parameter space, this mass kinematically forbids
 the equilibrating processes $\p\leftrightarrow \gamma\gamma$ if $\mds = m_\p < 2 \, m_t (T)$~\cite{Cadamuro:2010cz}.
However, for our fiducial choice of $\TXeq = 3\mds$, $m_\p > 2m_t(\TXeq)$ for any $m_\p$ and the results 
in Fig.~\ref{fig:neff_bounds_photon_equilibration} remain unaffected.
For other choices of equilibration temperatures and DS mass scales, this can be an important 
effect. For example, if we demand that $\TXeq = 5\mds$, then the decay channel is open only if $m_\ds < 100\;\keV$, 
significantly reducing the mass range over which $\p$ decays contribute to DS-photon equilibration.

Cosmologies in which a neutrino- or photon-coupled DS equilibrates after neutrino-photon decoupling demonstrate the need 
for both CMB- and BBN-based measurements of the expansion rate. 
Since these two epochs are widely separated in time, non-standard 
physics can affect one but not the other. 
This is particularly clear in the photon-coupled case, since the dominant 
change in the expansion rate can be localized in time between 
nucleosynthesis and recombination.
It is therefore 
crucial to constrain $\Delta \Neff$ at as many different cosmological times as possible. 
In specific models, additional constraints may be relevant. This is especially true 
for light, photon-coupled mediators, which we discuss in Appendix~\ref{sec:photon_constraints}.

\section{Conclusion}
\label{sec:conclusion}
  Measurements of the light element abundances provide a direct probe of the universe 
  seconds after the Big Bang. The concordance of the predictions 
  of standard Big Bang nucleosynthesis with observations of 
  \he4 and \D abundances constrains the existence of new physics 
  that contributes significantly to the energy density of the universe 
  at that time.
  We have investigated how a light dark sector that 
  comes into equilibrium with neutrinos or photons after neutrino-photon decoupling impacts the predictions of 
  \he4 and \D. This scenario naturally occurs if the processes 
  mediating energy exchange between the dark sector and the SM 
  become important at late times. 
  
  The equilibration and eventual decoupling 
  of new particles gives rise to a time-dependent modification of the 
  expansion rate, unlike the standard case of dark radiation.
  We have investigated several possibilities for this time-dependence, encoded in the temperature evolution of $\Neff$. 
  We considered both model-independent and concrete particle physics-motivated examples.
  Both approaches illustrate the seldom-mentioned point that, in general, $\Neff$ extracted from the CMB spectrum is different 
  from the one inferred from the light element abundances. Depending on the time-evolution of $\Neff$, 
  either the CMB or BBN can be the more sensitive probe of new physics.
  For example, if new degrees of freedom equilibrate with neutrinos after neutrino-photon decoupling, then the largest modification of the expansion rate occurs at late times. As a result, measurements of the CMB are typically more constraining for such processes. 
In fact, CMB-S4 experiments will decisively probe the relativistic neutrino-equilibration scenario, 
    which results in $\DNeff \gtrsim 0.2$ at the time of recombination~\cite{Berlin:2017ftj,Berlin:2018ztp,Abazajian:2016yjj}.
Instead, a dark sector that equilibrates with photons after neutrino-photon decoupling may dramatically alter the expansion rate during BBN without 
  significantly impacting it during recombination. In this case, observations of the light element abundances 
  provide the most important test of such new degrees of freedom. Regardless, CMB-S4 will achieve comparable sensitivity to these models in the coming decade.
These examples serve to illustrate the fact that measurements of the expansion rate during different epochs are crucial 
  in testing the viability of alternative cosmologies.

 In deriving the impact of a modified expansion history on the light element abundances, we have also highlighted the 
 importance of certain nuclear reactions. 
 Different choices for 
 key \D-burning rates significantly change both the central value of the prediction and the theoretical uncertainty, 
 leading to significantly different constraints on models of new physics. This ambiguity will be reduced with, e.g., the upcoming measurement of 
 $\D(p,\gamma)\he3$ at the LUNA experiment~\cite{Trezzi:2018qjs}.

 Our analysis relied on several simplifying assumptions. First, throughout this work we assumed that neutrinos have instantaneously decoupled from the baryon-photon 
 plasma, such that their entropy evolves independently for $T\lesssim 3\;\MeV$. In reality, the decoupling of electroweak interactions is gradual~\cite{Grohs:2016vef} and 
 neutrino-flavor dependent~\cite{Dolgov:2002wy}. It would be interesting to investigate how this non-instantaneous neutrino-decoupling affects constraints on models 
 in which dark sector equilibration or decoupling occurs near $T\sim \MeV$.
 As a first step, one can use the recent simplified neutrino decoupling method from Ref.~\cite{Escudero:2018mvt}. 
 Second, in the calculation of primordial abundances, we took the neutrino momentum distribution to be of the equilibrium 
 type. While this is a standard assumption in public BBN codes, new degrees of freedom will in general change the neutrino 
 spectrum and therefore affect certain 
 reaction rates. While the effect on the light nuclei yields from spectral distortions is expected to be small (e.g., in the SM, non-instantaneous 
 decoupling distorts neutrino distributions by $\lesssim 5\%$~\cite{deSalas:2016ztq}), it may become important as 
 the theoretical and observational uncertainties decrease.

 As mentioned above, light element abundances typically provide the strongest cosmological constraints on models where 
 new states equilibrate with photons after neutrino-photon decoupling. Such particles can also be produced in stellar environments, leading 
 to bounds from the observed lifetimes of horizontal branch and massive stars, and from the cooling 
 rate of SN1987A. We show in Appendix~\ref{sec:photon_constraints} that these bounds typically 
 exclude late photon equilibration cosmology. It is therefore important to understand whether this 
 statement is robust, or if there are simple models that avoid (or at least weaken) the stellar 
 constraints, through, e.g., environmental dependence of the dark sector-photon interactions.
 This is also interesting in the context of the proposed direct detection experiments 
 that seek to discover sub-MeV dark matter~\cite{Green:2017ybv,Knapen:2017xzo}.

\section*{Acknowledgment}
We thank Tongyan Lin, Sam McDermott, David Morrissey, Ken Nollett, and Josef Pradler for valuable discussions.
AB and SL are supported by the U.S. Department of Energy under Contract No. DE-AC02-76SF00515.
AB and NB thank the Kavli Institute of Theoretical Physics (KITP) where part of this work was completed. 
The research at KITP was supported in part by the National Science Foundation under Grant No. NSF PHY17-48958.
NB also thanks TRIUMF for hospitality during the completion of this work.
This manuscript has been authored by Fermi Research Alliance, LLC under 
Contract No. DE-AC02-07CH11359 with the U.S. Department of Energy, Office of Science, Office of High Energy Physics.

\appendix

\section{Collision Terms}
\label{sec:collision_terms}
In this section, we evaluate the collision terms responsible for 
the energy transfer between the visible and dark sectors. 
We focus on the simple example of two-body decays of a 
DS scalar, $\p$: $\p \leftrightarrow \nu\nu$ or $\p \leftrightarrow \gamma\gamma$,
which equilibrate the DS with neutrinos or with 
photons, respectively.
The energy transfer rate in and out of the $\p$ bath is then 
given by 
 \beq
\int \frac{d^3 p}{(2\pi)^3} C[f] = 
\int d\Phi_3 E_3 |\mathcal{M}|^2 (2\pi)^4 \delta^4(p_1 + p_2 - p_3) \left[f_1 f_2 (1 + f_3) - f_3 (1\mp f_1)(1\mp f_2)\right],
\label{eq:collision_term1}
\eeq
where $f_{1,2}$ and $f_3$ are distributions of the SM states ($\nu$ or $\gamma$) and 
of $\p$; $f_{1,2}$ is a function of $\Tsm$ (temperature of $\nu$ or $\gamma$), 
while $f_3$ depends on the DS temperature $\Tds$. In Eq.~\ref{eq:collision_term1}, the factors $(1\mp f_i)$ are 
Pauli-blocking or Bose-enhancement factors, with the $-$ sign relevant for 
$\p \to\nu\nu$ and $+$ for $\p \to \gamma\gamma$. The two-body decay 
matrix element $|\mathcal{M}|^2$ is related to the decay rate
\beq
|\mathcal{M}|^2 = 32\pi m_\p \Gamma_\p,
\eeq
where we took the final state particles to be massless. 
The $d\Pi_1 d\Pi_2$ phase-space integrals in Eq.~\ref{eq:collision_term1} can be 
performed by boosting into the $p_3$ rest-frame. The remaining integral 
over $p_3$ must be done numerically. The collision term can 
then be expressed as 
\beq
\int \frac{d^3 p}{(2\pi)^3} C[f] =
- m_\p \Gamma_\p \left[\Tds^3 h_1(r,x) - T^3_i h_2(r,x) \right],
\label{eq:collision_term2}
\eeq
where $r = \Tsm/\Tds$, $x = m_\p/\Tsm$. 
For $\p \rightarrow \nu\nu$ we find
\begin{align}
  h_1(r,x) & = \frac{r^3}{\pi^2}\int_x^\infty dy \left(\frac{y \exp(y)}{\exp(ry) - 1}\right)
  \frac{\ln \left(\cosh[(y+\sqrt{y^2 - x^2})/4]\sech[(y-\sqrt{y^2 - x^2})/4]\right)}{\exp y-1}\\
h_2(r,x) & = \frac{1}{\pi^2}\int_x^\infty dy \left(\frac{y \exp(r y)}{\exp(ry) - 1}\right)
  \frac{\ln \left(\cosh[(y+\sqrt{y^2 - x^2})/4]\sech[(y-\sqrt{y^2 - x^2})/4]\right)}{\exp y-1},
\end{align}
while the $\gamma\gamma$ final state gives the same expressions but with 
$\cosh \rightarrow \sinh$ and $\sech \rightarrow \csch$.
It is clear that since $h_1(1,x) = h_2(1,x)$, the collision term in Eq.~\ref{eq:collision_term2} vanishes as $\Tds\rightarrow \Tsm$. The $\gamma\gamma$ collision terms 
also exhibit Bose-enhancement in the form of a logarithmic singularity that is regulated by $m_\p$, resulting in a larger collision term and, therefore, 
faster equilibration for a given $\Gamma_\p$.

\section{Constraints on Dark Sector-Photon Equilibration}
\label{sec:photon_constraints}
In this section, we summarize the constraints on models 
of late DS-photon equilibration. As we showed in Sec.~\ref{sec:dark_sector_equilibration}, 
equilibration after neutrino-photon decoupling requires the presence of a light mediator particle 
with mass below the MeV scale. Such light states are easily produced in supernovae 
or in red giants and horizontal branch stars, if kinematically allowed. The broad agreement between the observed burst duration of 
SN1987A, the lifetimes of massive stars, and the corresponding SM predictions place tight constraints on the production of new 
particles that accelerate the energy loss in these systems~\cite{Raffelt:1996wa}. Here, we illustrate these difficulties in the context of an 
axion-like particle with the interaction
\beq
\mathscr{L}\supset \frac{\p}{4\Lambda} \widetilde{F}_{\mu\nu} F^{\mu\nu},
\eeq
where $\Lambda$ is a dimensionful scale related to the mass, coupling, and multiplicity of states of the ultraviolet physics that generates this interaction.
For example, if this operator is generated from a loop of heavy EM-charged states, then 
we expect $\Lambda \sim 2\pi M/\alpha_\text{em}$, where $\alpha_\text{em}$ is the fine-structure constant and $M$ is the 
mass of the heavy particles.

Thermalization of the visible and dark sectors is accomplished by decays and inverse decays, 
$\p \leftrightarrow \gamma\gamma$. The zero temperature rate is~\cite{Raffelt:2006cw} 
\beq
\Gamma_\p = \frac{m_\p^3}{64\pi\Lambda^2} .
\eeq
The visible and dark sectors equilibrate when $\Gamma_\p \times (m_\p/T) \sim H$ (see Sec.~\ref{sec:motivation}).
The equilibration temperature is then 
\beq
T^{\text{ds eq}} \sim 0.1 \left(\frac{m_\p \Mpl}{\Lambda^2}\right)^{1/3} m_\p. 
\eeq
If we conservatively demand that the equilibration and decoupling take place while 
SM photon number changing processes are efficient at $T \gtrsim 0.4\;\keV$~\cite{Chluba:2011hw} we find an approximate requirement on $\Lambda$
\footnote{Equilibration and decoupling may be viable at lower temperatures, since decoupling occurs in thermal equilibrium 
  and naively should not lead to CMB spectral distortions, as long as the DS particles have no chemical potential.}
\beq
\Lambda < 5\times 10^5 \;\GeV \times \left(\frac{m_\p}{\keV}\right)^2 \text{ (equilibration at $T \gtrsim 0.4\;\keV$)} .
\label{eq:lambda_upper_bound}
\eeq
These couplings are generically in conflict with limits from stellar cooling arguments. 
For example, for $m_\p < 10\;\keV$, the strongest constraint comes from the 
shortening of the He burning phase in Horizontal Branch (HB)~\cite{Ayala:2014pea} and massive stars~\cite{Friedland:2012hj}, leading to
\beq
\Lambda \gtrsim 10^{10}\; \GeV \text{ (stellar cooling)}.
\eeq
For $10 \text{ keV} \lesssim m_\p \lesssim 10 \text{ MeV}$, the dominant bound comes from observations of the SN1987A.
The agreement between the predicted and measured SM neutrino burst duration implies that~\cite{Lee:2018lcj}
\beq
\Lambda \gtrsim 10^{9}\;\GeV \text{ (SN1987A burst duration)}.
\eeq
If $\p$ decayed to photons once outside of the SN, even more powerful limits 
from the non-observation of a $\gamma$-ray excess become applicable~\cite{Jaeckel:2017tud}.
We see that generically the bounds on the $\p$ coupling to photons are in conflict with 
late equilibration.
While these limits are extremely powerful, they are not entirely model-independent.
For example, the $\p-\gamma$ interaction strength can be environment-dependent, and the effective 
coupling probed by the stellar cooling arguments above would not be the same as the 
coupling responsible for equilibration in the early universe~\cite{Brax:2007ak}. 
Cooling arguments can also be evaded via self-trapping inside of stars or supernovae~\cite{Jain:2005nh}.
Both environmental 
dependence of the $\p-\gamma$ coupling and self-trapping naturally occur if $\p$ has 
sufficiently strong self-interactions, i.e., a non-trivial potential.

\bibliographystyle{JHEP}
\bibliography{biblio}
\end{document}